\documentclass[11pt]{article}
\usepackage{amssymb,amsmath,fullpage,graphicx}

\def\sech{{\rm sech}}
\def\tanh{{\rm tanh}}
\def\csch{{\rm csch}}

\def\AllInt{ \displaystyle\int_{-\infty}^{\infty}}

\def\ri{{\rm i}}
\def\re{{\rm e}}
\def\rd{{\rm d}}
\def\Imag{\rm Imag}

\newcommand{\disp}{\displaystyle}

\begin{document}
\title{Exponential asymptotics for line solitons in
two-dimensional periodic potentials}
\author{Sean D. Nixon$^1$, T.~R.~Akylas$^2$, and Jianke Yang$^1$\footnote{e-mail:
jyang@math.uvm.edu} \\
{\small $^1$ Department of Mathematics and Statistics, University of
Vermont,
Burlington, VT 05401, USA} \\
{\small $^2$ Department of Mechanical Engineering, MIT, Cambridge,
MA 02139, USA}}
\date{ }
\maketitle

As a first step toward a fully two-dimensional asymptotic theory for
the bifurcation of solitons from infinitesimal continuous waves, an
analytical theory is presented for line solitons, whose envelope
varies only along one direction, in general two-dimensional periodic
potentials. For this two-dimensional problem, it is no longer viable
to rely on a certain recurrence relation for going beyond all orders
of the usual multi-scale perturbation expansion, a key step of the
exponential asymptotics procedure previously used for solitons in
one-dimensional problems. Instead, we propose a more direct
treatment which not only overcomes the recurrence-relation
limitation, but also simplifies the exponential asymptotics process.
Using this modified technique, we show that line solitons with any
rational line slopes bifurcate out from every Bloch-band edge; and
for each rational slope, two line-soliton families exist.
Furthermore, line solitons can bifurcate from interior points of
Bloch bands as well, but such line solitons exist only for a couple
of special line angles due to resonance with the Bloch bands. In
addition, we show that a countable set of multi-line-soliton bound
states can be constructed analytically. The analytical predictions
are compared with numerical results for both symmetric and
asymmetric potentials, and good agreement is obtained.

\bigskip

\section{Introduction}

Nonlinear wave propagation in periodic media is currently a subject
of intensive research in optics and applied mathematics, with
diverse applications ranging from nonlinear photonics
\cite{Kivshar2003,Skorobogatiy2009} to Bose--Einstein condensates
\cite{Morsch2006,Kevrekidis2008}. While many types of soliton
structures in periodic media have been reported theoretically and
experimentally (see \cite{Yang_SIAM} for a review), prior analytical
work is limited to special (symmetric) periodic potentials
\cite{Dohnal2009, Ilan2010}, or to one spatial dimension
\cite{Hwang2011, Akylas2012, Hwang2012}.

In this paper, we study line solitons in general two-dimensional
periodic media, employing the nonlinear Schr\"odinger (NLS) equation
with a spatially periodic potential as our model. Such solitons have
been reported theoretically and experimentally in certain periodic
potentials before
\cite{Chen2004,Lou2007,Wang2007,Yang2011,PeliYang2012}, and it is
known that they can bifurcate from edges of Bloch bands or from
certain high-symmetry points inside Bloch bands. However, systematic
analytical construction of line solitons is still lacking, and a
number of open questions remain. It is not clear, in particular, if
the inclination angles and positions of these line solitons relative
to the underlying potential can be arbitrary or not.

To address these issues, in this article we analytically examine
small-amplitude line solitons which bifurcate out from infinitesimal
Bloch modes. Such solitons are in the form of slowly varying
Bloch-wave packets, with the envelope of the packet being uniform
along the line direction. Using standard asymptotic methods, the
packet envelope can be readily shown to satisfy the familiar
one-dimensional NLS equation in the absence of a potential and thus
have a sech-shape. However, the position of the envelope relative to
the potential is harder to determine because it hinges on effects
that are exponentially small in the soliton amplitude. For this
purpose, techniques of exponential asymptotics must be used.

In one-dimensional nonlinear wave systems, an exponential
asymptotics method for analyzing low-amplitude solitons that
comprise a periodic carrier modulated by an envelope, has been
developed in the past fifteen years
\cite{Hwang2011,Akylas2012,Hwang2012,Akylas1997}. This method
focuses on the Fourier transform with respect to the slow spatial
variable of the envelope function, motivated by the fact that the
solitary-wave tails in the physical domain are controlled by the
pole singularities near the real axis of the wavenumber space. The
residues of these poles, which are exponentially small, are then
calculated by matched asymptotics near the poles and away from the
poles. The solution away from the poles, in particular, is expressed
as a Taylor series of the wavenumber around zero, multiplied by a
factor which is exponentially small at the poles, and the
coefficients in this series are computed through a recurrence
relation. Upon inverting the Fourier transform, these poles of
exponentially small strength give rise to growing tails of
exponentially small amplitudes in the physical solution. It turns
out that these growing tails vanish only at two positions of the
envelope relative to the periodic carrier, which in turn reveals
that only two soliton families are admitted.

In this paper, by investigating line solitons, we present the first
step in developing a fully two-dimensional asymptotic theory for the
bifurcation of solitons from infinitesimal continuous waves. For
line solitons, the governing equation for the carrier envelope is
still one-dimensional, hence the corresponding Fourier transform
depends on only one complex variable in the wavenumber domain. While
this simplifies matters, two new obstacles arise. The first is that,
in a two-dimensional problem, the Fourier transform of the solution
contains pole singularities not only near the real axis (which we
call real poles), but also away from the real axis (which we call
complex poles). Which of these poles are relevant for the
determination of line solitons is unclear. On this issue, we will
show that it is still the real poles which matter.

The second and more serious obstacle is that, in two dimensions, the
poles closest to the origin are often complex rather than real. As a
consequence, utilizing a recurrence relation to determine the
residues of real poles, a key step in one-dimensional problems, is
no longer viable: the radius of convergence of the Taylor series
(around the zero wavenumber) is limited by the nearest complex poles
and thus cannot reach the real poles further away, which makes the
matched asymptotics impossible to use. To overcome this obstacle, we
shall give up the recurrence-relation approach and directly solve
the Fourier-transformed equation along the real line of the
wavenumber plane, up to the real poles. By doing so, complex poles
become irrelevant and thus can be ignored. This modified approach
not only turns out to be effective for two dimensions, but also
simplifies the exponential asymptotics procedure: by working with
the Fourier-transformed equation, the exponentially-small terms are
factored out and one is left with a Volterra integral equation,
which is in fact easier to solve numerically than the recurrence
relation.

We apply this modified exponential asymptotics technique to line
solitons bifurcating from edges of Bloch bands in general
two-dimensional periodic potentials. At low amplitudes, the solution
is a Bloch-wave packet whose envelope varies only in the direction
normal to the line soliton. By using the Fourier-transform approach
as outlined above, we analytically calculate growing tails of
exponentially small amplitudes in the normal direction. The results
show that for any rational slope, two line solitons relative to the
potential exist, and the positions of their envelopes can be
explicitly obtained. By matching the growing and decaying tails, we
also construct an infinite family of multi-line-soliton bound
states. These line solitons and multi-line-soliton bound states have
been obtained numerically as well, in agreement with the theory. To
quantitatively verify the analytical formula for the growing tails
of exponentially-small amplitude, we use this tail formula to
calculate the bifurcation of the zero eigenvalue associated with the
linear stability of low-amplitude line solitons. This analytical
eigenvalue formula is compared with the numerical eigenvalues, and
good quantitative agreement is reached. Finally, we also show that
line solitons can bifurcate from interior high-symmetry points of
Bloch bands as well. But such line solitons exist only for very few
(up to three) special line angles due to resonance with the Bloch
bands.

\section{Line wavepackets at low amplitudes}
\label{Sec: LineSoliton}

We consider the nonlinear Schr\"{o}dinger equation in two spatial
dimensions with a periodic (lattice) potential,
\begin{equation}
\ri \Psi_t + \nabla^2 \Psi - V(x,y)\Psi + \sigma |\Psi |^2 \Psi = 0,
\label{Eq: NLS}
\end{equation}
where $\nabla^2=\partial_{xx}+\partial_{yy}$, the potential $V(x,y)$
is periodic in both $x$ and $y$, and $\sigma=\pm 1$ is the sign of
nonlinearity. This equation governs the nonlinear propagation of
light in photonic lattices as well as the evolution of
Bose--Einstein condensates in optical lattices
\cite{Kivshar2003,Skorobogatiy2009, Morsch2006, Pitaevskii2003}. In
the latter community, this equation is referred to as the
Gross$-$Pitaevskii equation.

In this article, we shall take the minimal periods of the potential
$V(x,y)$ to be the same in both $x$ and $y$. This matching
periodicity in $x$ and $y$, while not necessary for our analysis,
allows for simplification of the algebra. For a periodic potential
with matching periodicity, rotation by an angle $\theta$, with
$\tan(\theta)$ rational, also yields a periodic potential, but the
periods of the new potential would change in general. For instance,
for the potential $V(x,y)=\sin^2x+\sin^2y$ with matching periods
$\pi$, a $45^\circ$-rotation would yield a potential with matching
periods $\sqrt{2}\hspace{0.02cm}\pi$. To remove this rotational
freedom and the changing periods of the same potential, we align the
potential in such a way that its periods are minimal. We call the
potential in such orientation as the minimal-period-orientation
potential, and this potential is used throughout the paper. Without
loss of generality, we also normalize the matching periods of this
minimal-period-orientation potential as $\pi$.

We search for stationary solutions to Eq. (\ref{Eq: NLS}) of the
form
\begin{equation}
\Psi(x,y,t) = \psi(x,y) \re^{-\ri \mu t},
\end{equation}
where $\mu$ is the propagation constant and $\psi$ is a real-valued
amplitude function that satisfies the equation
\begin{equation}
\nabla^2 \psi + \left( \mu - V(x,y) \right) \psi + \sigma \psi^3 = 0.
\label{Eq: psi}
\end{equation}

In this article, we shall consider line-soliton solutions that are
bounded (non-decaying) along a certain line in the $(x,y)$ plane but
decay to zero along the direction orthogonal to this line. These
solutions may bifurcate out from edges of Bloch bands, or from
certain high-symmetry points inside Bloch bands
\cite{Lou2007,Wang2007,Yang2011,PeliYang2012}. Our analysis will be
developed first for line solitons bifurcating from edges of Bloch
bands. Afterwards (in section \ref{Sec: Xpoint}) we shall consider
line solitons bifurcating from interior points of the Bloch bands,
which do exist but under more restrictive conditions due to
resonance with Bloch bands.

Near edges of Bloch bands, solitons are low-amplitude wavepackets
comprising the underlying Bloch-wave carrier modulated by a slowly
varying envelope, and they can be studied using multiple-scale
perturbation theory. For line solitons, the envelope only varies in
one direction and the associated `slow' spatial variable will be
\begin{equation}  \label{Eq: Wdef}
W =\epsilon w, \quad w\equiv x\sin\theta -y\cos\theta,
\end{equation}
where $\theta$ is the angle the line soliton makes with the $x$
axis, and $\sqrt{|\mu -\mu_0|} = \epsilon \ll 1$ is the deviation
from the edge point $\mu_0$. Plugging this into equation \eqref{Eq:
psi} yields
\begin{equation}
L_0 \psi +  \epsilon \partial_W L_1 \psi  + \epsilon^2 \partial_W^2
\psi + \sigma \psi^3 + (\mu - \mu_0) \psi = 0, \label{Eq:
MultiScale}
\end{equation}
where
\begin{equation}
L_0 = \nabla^2 + \mu_0 - V(x,y), \quad  L_1 = 2 \nabla \cdot \left[
\sin\theta, -\cos\theta \right].
\end{equation}
The perturbation series solution to Eq. (\ref{Eq: MultiScale}) is
readily found to be
\begin{equation}
\psi(x,y,W) = \epsilon A(W) b(x,y) + \epsilon^2 A'(W) \nu(x,y) +
O(\epsilon^3),  \label{Eq: Aseries}
\end{equation}
where the functions of the fast variables satisfy
\begin{eqnarray}
L_0 b(x,y) &=& 0,   \label{E: L0p} \\
L_0 \nu(x,y) &=& - L_1 b(x,y).   \label{E: L0nu}
\end{eqnarray}
Eq. (\ref{E: L0p}) implies that $b(x,y)$ is a Bloch mode at the edge
point $\mu=\mu_0$. For simplicity of the analysis, we shall make the
following assumption.

\vspace{0.2cm} \textbf{Assumption 1} \ The Bloch mode at the band
edge $\mu=\mu_0$ is unique (up to a multiplication constant).

\vspace{0.2cm} This assumption is for the purpose of avoiding
nonlinear interactions between different Bloch modes at the same
band-edge point, and it holds for many band edges. We also normalize
$b(x,y)$ so that ${\rm max} |b|=1$.

The inhomogeneous equation (\ref{E: L0nu}) is solvable, since the
right-hand side $-L_1b$ is clearly orthogonal to the homogeneous
solution $b(x,y)$ so the Fredholm condition is satisfied. To avoid
ambiguity of the homogeneous term in $\nu$, we require $\langle
\nu,b\rangle = 0$, where the inner product is defined as
\begin{equation}
\langle f, g\rangle=\int_0^{2\pi} \int_0^{2\pi}  f^*(x,y)g(x,y) \
\rd x \rd y,
\end{equation}
with the superscript `*' representing complex conjugation.

From the solvability condition at order $\epsilon^3$, the envelope
function $A(W)$ in expansion (\ref{Eq: Aseries}) satisfies
\begin{equation}
D A''+ \eta A  + \sigma a A^3 = 0,  \label{Eq: Envelope}
\end{equation}
with
\begin{equation}
D = \frac{\langle L_1 \nu + b , b\rangle}{ \langle b, b\rangle },
\quad \eta=\mbox{sgn}(\mu-\mu_0),  \quad a = \frac{\langle b^3,
b\rangle }{\langle b, b\rangle }. \label{Eq: aD}
\end{equation}
When sgn($\sigma$) = sgn($D$) = $-$sgn($\eta$), the solution of the
envelope equation \eqref{Eq: Envelope} is
\begin{equation}
A(W) = \alpha \hspace{0.06cm} \sech \frac{W - W_0}{\beta},
\label{Eq: A}
\end{equation}
where $W=W_0$ is the center position of the line envelope, and
\begin{equation}
\alpha = \sqrt{2/a}, \quad \beta = \sqrt{|D|}.
\end{equation}

\begin{figure}[htbp!]
\begin{center}
\includegraphics[width=4.9in]{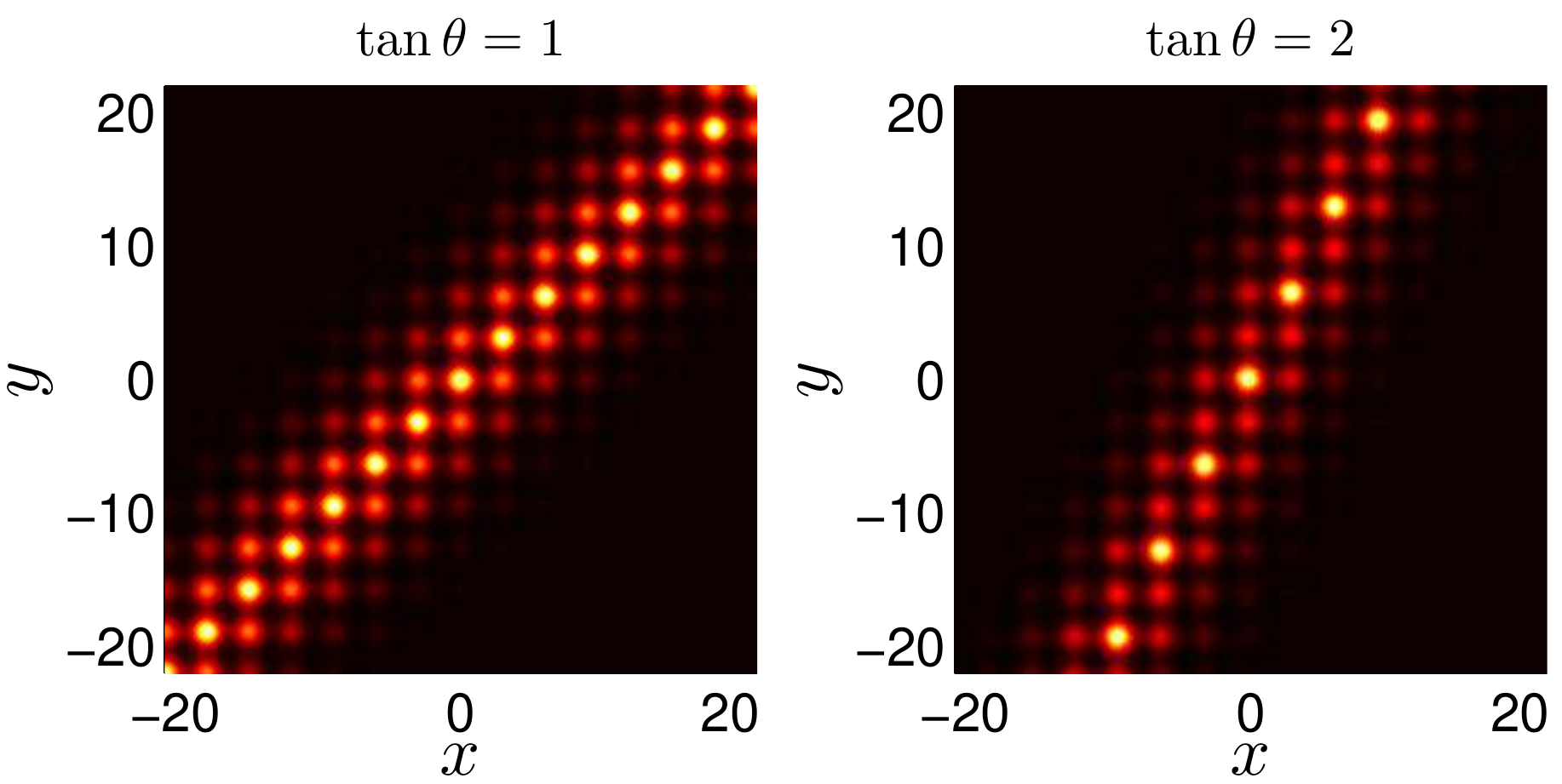}
\includegraphics[width=4.9in]{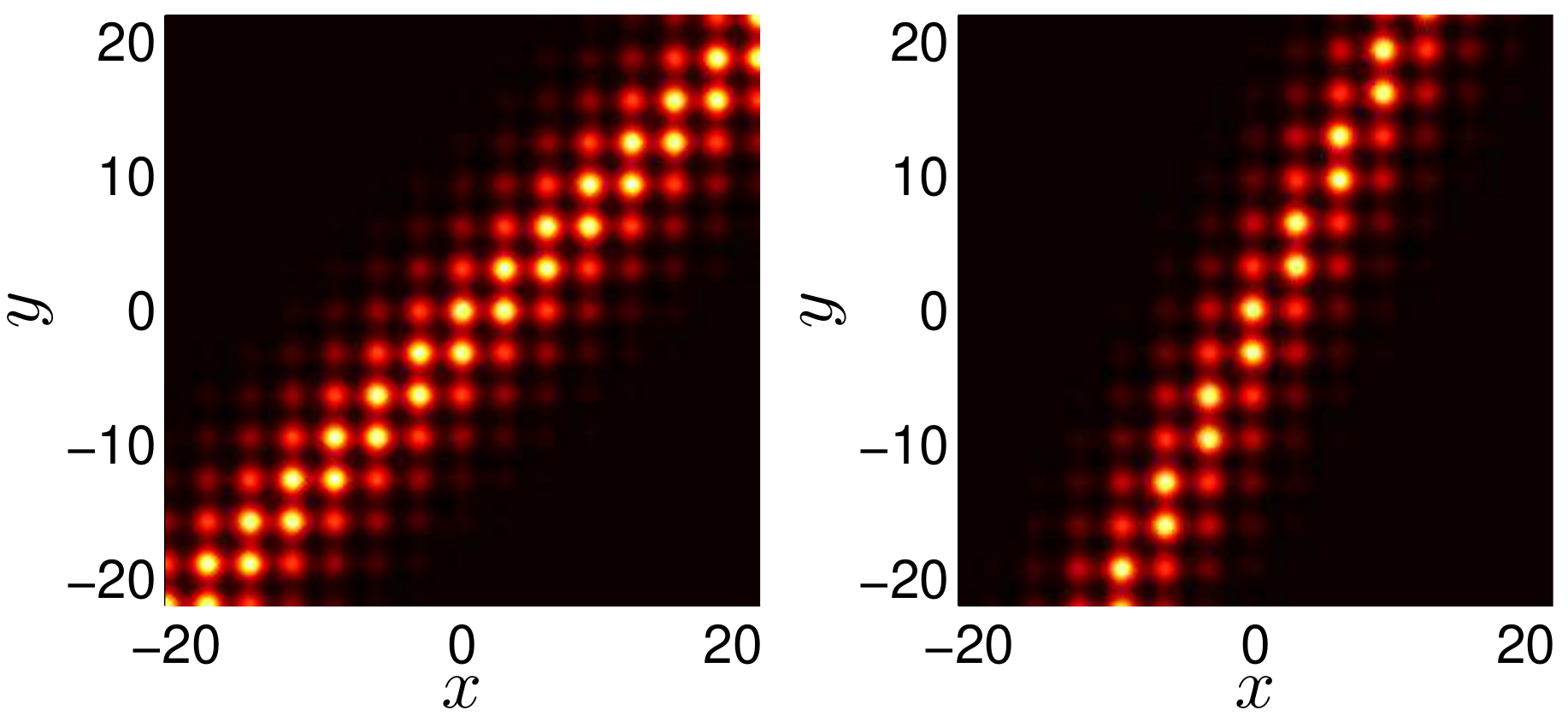}
\caption{(Color online) Line solitons for $\sigma=1$, $\epsilon =
0.25$ and $\tan\theta = 1,2$ in the lattice (\ref{Eq: ExLat}). Upper
row: onsite solitons; lower row: offsite solitons.
\label{Fig:LSolitons}}
\end{center}
\end{figure}

The asymptotic expansion \eqref{Eq: Aseries} may be carried to all
powers of $\epsilon$, with all terms being localized in $W$ (i.e.,
localized along the normal direction of the line solution),
suggesting that line solitons exist for all choices of $W_0$.
However, based on previous experience and numerical computations,
this is not the case. Specifically, it turns out that for rational
$\tan \theta$, only two line-soliton families exist (not counting
their periodic replications in the lattice). These two soliton
families are illustrated in Fig. \ref{Fig:LSolitons} for $\sigma=1$,
$\epsilon=0.25$ and $\tan\theta = 1,2$ in the specific lattice
\begin{equation}
V(x,y) = 6\left( \sin^2\hspace{-0.05cm}x + \sin^2\hspace{-0.05cm} y
\right), \label{Eq: ExLat}
\end{equation}
where they bifurcate out from the lowest Bloch-band edge
$\mu_0=4.1264$. Solitons in the upper panels have $W_0=0$ and are
called onsite solitons, while those in the lower panels have
$W_0=\epsilon \pi/2$ and are called offsite solitons.

This discrepancy between the prediction of the multi-scale expansion
(\ref{Eq: Aseries}) and true solutions was first studied in
\cite{Pelinovsky2004} for a one-dimensional problem, and it was
suggested that the issue could be resolved by requiring that a
Melnikov-integral condition be satisfied by the leading-order
approximation of \eqref{Eq: Aseries}. While this constraint happens
to specify $W_0$ correctly for symmetric potentials, in a general
lattice all terms in the perturbation expansion \eqref{Eq: Aseries}
make contributions to the Melnikov integral at the same order of
$\epsilon$, and the calculations involved quickly get out of hand.
To handle this difficulty, an exponential-asymptotics approach was
developed for one-dimensional problems in
\cite{Hwang2011,Hwang2012}, following an earlier treatment of a
similar problem in the fifth-order KdV equation \cite{Akylas1997}.
In the next sections, we shall adapt this exponential-asymptotics
method to the study of line solitons. For this two-dimensional
problem, certain key steps in the previous exponential-asymptotics
procedure are no longer viable, and suitable modifications will be
needed in order to overcome those obstacles.

\section{Solution in the wavenumber domain}
\label{Sec: FourierDomain}

The failure of the multi-scale expansion \eqref{Eq: Aseries} stems
from the fact that, for generic values of the envelope position
$W_0$, the exact solution $\psi(x,y;W)$ contains growing tails when
$W\ll -1$ or $W \gg 1$; but the amplitudes of these tails are
exponentially small in $\epsilon$, and hence invisible in the
expansion \eqref{Eq: Aseries}. In order to obtain true line
solitons, we shall first calculate these exponentially small growing
tails and then determine the envelope position $W_0$ by insisting
that these tails vanish.

As before \cite{Hwang2011,Hwang2012, Akylas1997}, we shall work in
the wavenumber domain. First, we take the Fourier transform of
$\psi(x,y,W)$ with respect to the slow variable $W$,
\begin{equation}  \label{Eq: FT}
\widehat{\psi}(x,y,K) = \frac{1}{2\pi} \disp\int_{-\infty}^{\infty}
\psi(x,y,W) \re^{-\ri KW} \rd W.
\end{equation}
Under this transformation, the growing tails of exponentially small
amplitude in $\psi(x,y,W)$ are transformed into poles of
$\widehat{\psi}$ with exponentially small residues.

Taking the Fourier transform of the series solution \eqref{Eq:
Aseries} term-by-term yields
\begin{equation}
\widehat{\psi}(x,y,K) = \epsilon \frac{\alpha \beta}{2}   \re^{-\ri
W_0 K} \sech \left(\frac{\pi\beta K}{2} \right) \left\{ b(x,y) + \ri
\epsilon K \nu(x,y) + \ldots \right\}. \label{Eq: psihat}
\end{equation}
This series is disordered at $\epsilon K\sim 1$. Thus, we introduce
the `slow' wavenumber $\kappa = \epsilon K$ and rearrange this
series as
\begin{align}
\widehat{\psi}(x,y,\kappa)=\epsilon \hspace{0.05cm} \re^{-\ri W_0
\kappa /\epsilon} \sech \left(\frac{\pi\beta\kappa}{2\epsilon}
\right) U(x,y,\kappa; \epsilon),  \label{Eq: Psihat}
\end{align}
where
\begin{equation} \label{Eq: Useries}
U(x,y,\kappa; \epsilon) = \frac{\alpha \beta}{2} \left\{b(x,y) + \ri
\kappa \nu(x,y) + \ldots\right\}, \quad \kappa \ll 1.
\end{equation}

We now derive the governing equation for $U$ by taking the Fourier transform of equation \eqref{Eq: psi} to arrive at
\begin{equation}
L_0 \widehat{\psi} + \ri \kappa L_1 \widehat{\psi}   - \kappa^2
\widehat{\psi} + \sigma \widehat{\psi^3} + \epsilon^2 \eta
\widehat{\psi} = 0.
\end{equation}
Then, by substituting the expression \eqref{Eq: Psihat}, we find
that $U$ satisfies
\begin{align}
\label{Eq: UFull} L_0 U &+ \ri \kappa L_1 U + (\eta\epsilon^2 -
\kappa^2) U + \sigma
\cosh\left( \frac{\pi \beta \kappa}{2 \epsilon} \right) \times   \nonumber \\
&\AllInt \AllInt \frac{ U(\kappa- r) U(r - s) U(s)} { \cosh\left(
\frac{\pi \beta (\kappa- r)}{2 \epsilon} \right) \cosh\left(
\frac{\pi \beta (r - s)}{2 \epsilon} \right) \cosh\left( \frac{\pi
\beta s}{2 \epsilon} \right)} \rd r \rd s=0.
\end{align}

\section{Poles in the wavenumber plane}
\label{Sec: Poles}

We are concerned with pole singularities in $U(x,y,\kappa;
\epsilon)$ which account for the growing tails in the physical
space. Singularities of $U$ are expected to occur near values of
$\kappa=\kappa_0$ where the linear part of equation \eqref{Eq:
UFull} is zero, i.e.,
\begin{equation}
L_0 \phi + \ri \kappa_0 L_1 \phi  - \kappa_0^2 \phi= 0.  \label{Eq:
LinearU}
\end{equation}
With a change of variables $\phi= \re^{-\ri \kappa_0 w}
\widetilde{\phi}$, where $w$ is defined in Eq. (\ref{Eq: Wdef}),
equation (\ref{Eq: LinearU}) reduces to
\begin{equation} \label{Eq: L0tildephi}
L_0 \widetilde{\phi} = 0,
\end{equation}
which has a single bounded solution $\widetilde{\phi} = b(x,y)$, the
Bloch mode at band edge $\mu_0$. Thus, if we restrict ourselves for
the moment to real values of $\kappa_0$, bounded solutions to Eq.
(\ref{Eq: LinearU}) are
\begin{equation}
\phi(x,y)=\re^{-\ri \kappa_0 w} b(x,y).
\end{equation}
Since the spatial period of the solution $\phi(x,y)$ should match
that of the solution (\ref{Eq: Useries}), $\phi(x,y)$ and $b(x,y)$
should have the same periodicity in $(x,y)$. Then following the
argument in \cite{Hwang2011}, both $\kappa_0 \cos(\theta)$ and
$\kappa_0 \sin(\theta)$ should be even integers. For this to occur,
$\tan\theta$ must be a rational number, say
\begin{equation}
\tan(\theta) =p/q,
\end{equation}
where $p$ and $q$ are relatively prime. Now to satisfy the
periodicity condition, we get
\begin{equation} \label{Eq: realkappa0}
\kappa_0 = 2n \sqrt{p^2+q^2}
\end{equation}
for any integer $n$. Thus poles of $U(x,y,\kappa; \epsilon)$ near
the real axis of $\kappa$ are located near these $\kappa_0$ values.

To get the approximate locations of all poles in $U(x,y,\kappa;
\epsilon)$, we repose Eq. (\ref{Eq: LinearU}) as an eigenvalue
problem
\begin{equation}
\ri \begin{bmatrix} L_1& L_0 \\ -1& 0 \end{bmatrix} \vec{\phi} =
\kappa_0 \vec{\phi}, \label{Eq: EigenProblem}
\end{equation}
where $\vec{\phi} = [\phi, ~(1/\ri\kappa_0)\phi]^T$, with the
superscript `$T$' denoting vector transpose. It can be easily shown
that all eigenvalues in (\ref{Eq: EigenProblem}) come in quadruples
($\kappa_0,-\kappa_0,\kappa_0^*,-\kappa_0^*$). This spectrum also
has periodicity $2\sqrt{p^2 + q^2}$ in the real direction of
$\kappa_0$.

\begin{figure}[htbp!]
\begin{center}
\includegraphics[width=4.9in, height=2.4in]{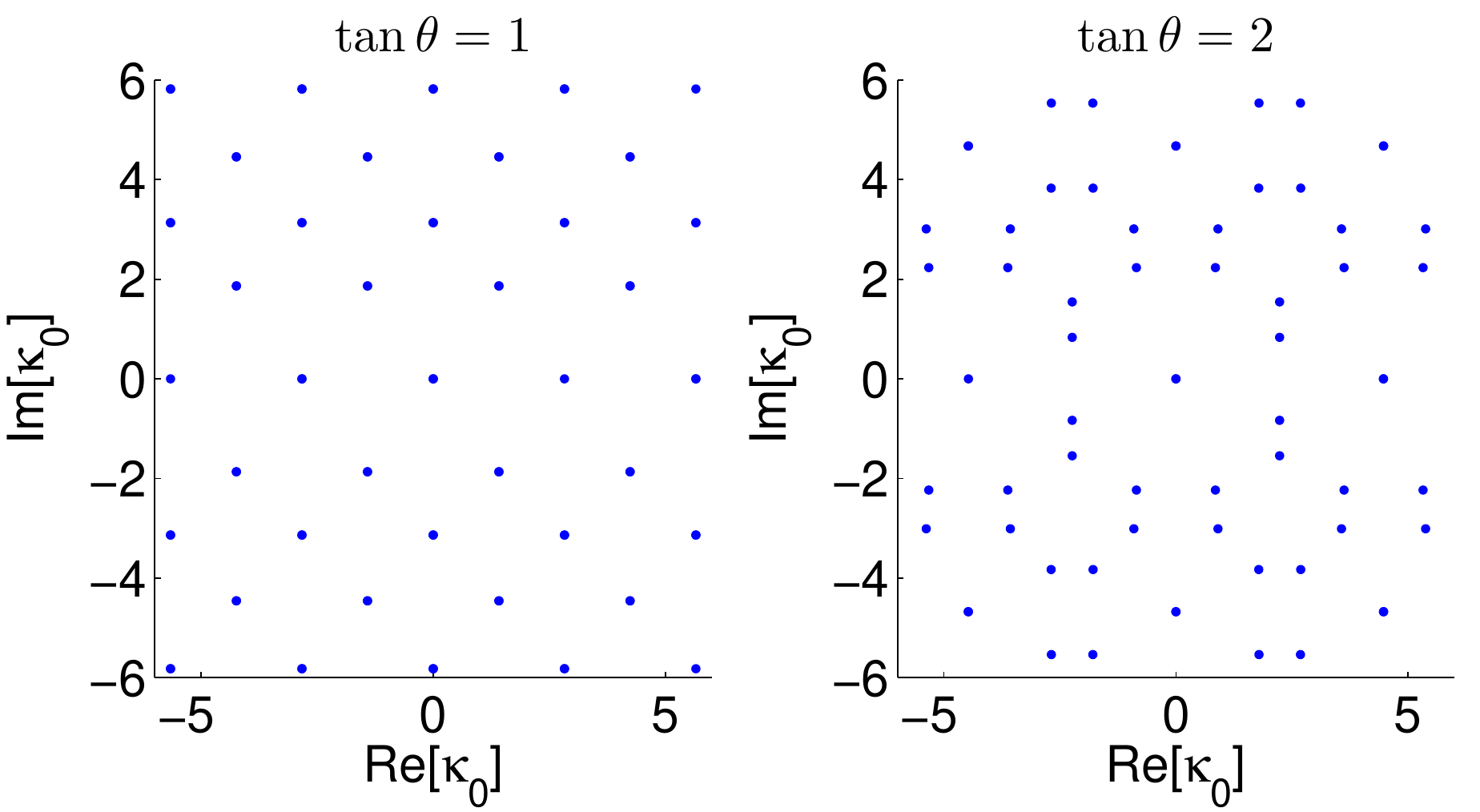}
\caption{Locations of singularities in the wavenumber domain for the
lattice (\ref{Eq: ExLat}) with $\tan\theta = 1,2$.
\label{Fig: LSpecTan12} }
\end{center}
\end{figure}

Numerically we solve the eigenvalue problem (\ref{Eq: EigenProblem})
using the Fourier collocation method to get rough estimates of the
spectrum, followed by the Newton-conjugate-gradient method to
calculate particular eigenvalues to high accuracy \cite{Yang_SIAM}.
Examples of this spectrum are shown in Fig. \ref{Fig: LSpecTan12}
for the lattice \eqref{Eq: ExLat} at $\mu_0=4.1264$ (the edge of the
semi-infinite gap) and $\tan\theta=1,2$. Notice that this spectrum
contains not only the real eigenvalues given by Eq. (\ref{Eq:
realkappa0}), but also a large number of complex eigenvalues. In
addition, when $\tan\theta \neq 0$, the eigenvalues closest to the
origin are complex eigenvalues rather than real eigenvalues.

The spectra in Fig. \ref{Fig: LSpecTan12} raise serious questions on
the applicability of the exponential-asymptotics method, as used
earlier for solitons in one-dimensional lattices, to this
two-dimensional problem. We recall that, in the one-dimensional case
\cite{Hwang2011,Hwang2012,Akylas1997}, this spectrum contains only
real eigenvalues; and it is those real eigenvalues which specify the
envelope positions. In the present case, where both real and complex
eigenvalues exist, which of those eigenvalues are responsible for
selecting the envelope positions?

To tackle this issue, we consider line solitons in a simpler stripe
lattice, where the lattice varies in the $x$-direction only. The
particulars of the analysis are covered in the appendix. For such a
lattice, we find that the spectrum contains only complex eigenvalues
and no nonzero real eigenvalues. Here, however, the envelope
position of a line soliton can be arbitrary, as solutions with
different envelope positions are equivalent to each other under a
vertical translation. This indicates that the envelope position is
not affected by complex eigenvalues, and thus we shall focus on
poles of $U(x,y,\kappa; \epsilon)$ near the real eigenvalues
(\ref{Eq: realkappa0}). In particular, we need to determine the
residues of poles near the smallest real eigenvalues
\begin{equation}  \label{Eq: kappa0}
\kappa_0 = \pm 2 \sqrt{p^2+q^2},
\end{equation}
since those poles provide the dominant contributions to the
solitary-wave tails.

Even though only real eigenvalues affect the envelope position, the
fact that the eigenvalues closest to the origin are complex rather
than real, as shown in Fig. \ref{Fig: LSpecTan12}, poses a major
obstacle to the calculation of the residues of poles near the real
eigenvalues $\kappa_0 = \pm 2 \sqrt{p^2+q^2}$ by the previous
exponential-asymptotics technique \cite{Hwang2011, Hwang2012,
Akylas1997}. This difficulty and its resolution will be elaborated
in the next section.

\section{Solution away from the poles}
\label{Sec: Near0}

The basic idea for calculating the residues of poles in
$U(x,y,\kappa; \epsilon)$ is to match the `inner' solution near the
poles to the `outer' solution away from the poles. Since the poles
of interest are near $\kappa_0 =\pm 2\sqrt{p^2 + q^2}$, in this
section we will determine the solution $U(x,y,\kappa; \epsilon)$ for
real values of $\kappa$ in the interval $-\kappa_0 < \kappa <
\kappa_0$ but not close to $\pm \kappa_0$.

When $\epsilon \rightarrow 0$, the main contribution of the double
integral in \eqref{Eq: UFull} comes from the triangular region $0< s
<\kappa$, $0< s < r$ when $\kappa
>0$ or $\kappa<s<0$, $r<s<0$ when $\kappa < 0$. Over this region,
\begin{equation}
\cosh\left( \frac{\pi \beta \kappa}{2 \epsilon} \right) \left /
\left[\cosh\left( \frac{\pi \beta (\kappa- r)}{2 \epsilon} \right)
\cosh\left( \frac{\pi \beta (r - s)}{2 \epsilon} \right) \cosh\left(
\frac{\pi \beta s}{2 \epsilon} \right)\right] \approx 4,  \right.
 \label{Eq: NLRecursion}
 \end{equation}
while outside this region the same expression is exponentially
small. Thus, to $O(\epsilon^2)$ the integral equation \eqref{Eq:
UFull} reduces to the ``outer" equation
\begin{align}
L_0 U^{(0)} &+ \ri \kappa L_1 U^{(0)}   -  \kappa^2 U^{(0)} \notag \\
 & \hspace{-0.8cm} + 4\sigma \disp\int_0^{\kappa} \rd r \ U^{(0)}
 (x,y,\kappa- r) \disp\int_0^r \rd s  \ U^{(0)} (x,y,r - s) U^{(0)} (x,y,s)=0,
\label{Eq: Uorder1}
\end{align}
where $U(x,y,\kappa; \epsilon) =  U^{(0)}(x,y,\kappa) +
O(\epsilon^2)$. Here the main contribution to the error of the
integral comes from the three corners of the triangular integration
region in the integral of (\ref{Eq: Uorder1}); thus this error is
$O(\epsilon^2)$, a result that has also been checked numerically for
randomly chosen analytic functions of $U(\kappa)$.

In previous applications of exponential asymptotics, this outer
integral equation was solved by expanding $U^{(0)}(x,y,\kappa)$ into
a power series of $\kappa$, which turns the outer integral equation
into a recurrence relation for the coefficients of the power series
\cite{Hwang2011, Hwang2012, Akylas1997}. This treatment was possible
since the poles of interest were closest to the origin, hence the
power series was convergent up to those poles. In the present
two-dimensional problem, however, this treatment fails, because the
power series for $U^{(0)}(x,y,\kappa)$ has radius of convergence
bounded by the distance to the nearest complex poles. Thus this
power series cannot converge near the real poles of interest. As a
result, the previous approach of relying on the recurrence relation
needs to be abandoned.

Instead, we propose to solve the outer integral equation (\ref{Eq:
Uorder1}) numerically. Since this is a Volterra integral equation,
it can be easily tackled by explicit numerical methods. First we
discretize $\kappa$, $\kappa_n = n \Delta \kappa$, and write
\begin{equation}
U^{(0)}(x,y, \kappa_n) = U_n(x,y).
\end{equation}
Then we approximate the integral in (\ref{Eq: Uorder1}) using the
trapezoid rule. After the terms in the resulting equation are
rearranged, $U_n$ is found to satisfy a linear inhomogeneous
equation
\begin{equation}
\left[L_0  + \ri\kappa_n L_1 - \kappa_n^2 + 2\sigma U_0^2 (\Delta
\kappa)^2 \right] U_n  = - 4\sigma F_n, \label{Eq: Iteration}
\end{equation}
where the inhomogeneous term $F_n$ is given by
\begin{subequations}
\begin{align}
F_n &= \Delta \kappa^2 \left[ \frac{1}{2} U_0 I_n + \displaystyle
\sum_{m=1}^{n-1} U_m I_{n-m}\right], \\
I_m &=\displaystyle\sum_{l = 0}^{m-1}U_l U_{m-l}, \quad {\rm for}~ 1\le m<n, \\
I_n &= \displaystyle\sum_{l = 1}^{n-1}U_l U_{n-l}.
\end{align}
\end{subequations}
Since the inner sums $I_m$ for $m<n$ do not change on further
iterations, these only need to be computed once. Notice that the
homogeneous operator in Eq. \eqref{Eq: Iteration} is self-adjoint,
thus this linear inhomogeneous equation can be solved by the
preconditioned conjugate gradient method. The initial conditions
$U_0$ and $U_1$ cannot be derived from Eq. \eqref{Eq: Iteration}
itself, but they can be obtained from the equation (\ref{Eq:
Useries}) as
\begin{equation}
U_0(x,y) =\frac{\alpha \beta}{2} b(x,y),  \quad U_1(x,y) =
\frac{\alpha \beta}{2}\left[b(x,y) + \ri \Delta \kappa~
\nu(x,y)\right].
\end{equation}

The above numerical scheme for solving the outer integral equation
(\ref{Eq: Uorder1}) is explicit. Our numerical testing shows that
its numerical error is $O(\Delta \kappa)$, thus it is first order
accurate in $\Delta \kappa$. If one wishes for a higher-order
numerical scheme, then instead of the trapezoidal rule, one can use
a higher-order quadrature method (such as Simpson's rule) to
approximate the integral in (\ref{Eq: Uorder1}).

From the local analysis near the poles in the next section, it will
transpire that $U^{(0)}(x,y,\kappa)$ has a fourth-order pole at
$\kappa_0$. Specifically,
\begin{equation}
U^{(0)}(x,y,\kappa) \to \frac{12 C}{5} \frac{b(x,y)}{(\kappa -
\kappa_0)^4}\re^{-\ri \kappa_0 w},  \quad \mbox{as} \ \kappa \to
\kappa_0, \label{Eq: UAsymptotics}
\end{equation}
where $C$ is a complex constant. With a change of variables
\begin{equation}
\widetilde{U}(x,y, \kappa) = (\kappa-\kappa_0)^{4}
U^{(0)}(x,y,\kappa),
\end{equation}
then
\begin{equation}
\widetilde{U}(x,y,\kappa) \to \frac{12 C}{5} b(x,y) \re^{-\ri
\kappa_0 w},  \quad \mbox{as} \ \kappa \to \kappa_0. \label{Eq:
UtildeAsymptotics}
\end{equation}

\begin{figure}[htbp!]
\begin{center}
\includegraphics[width=5in]{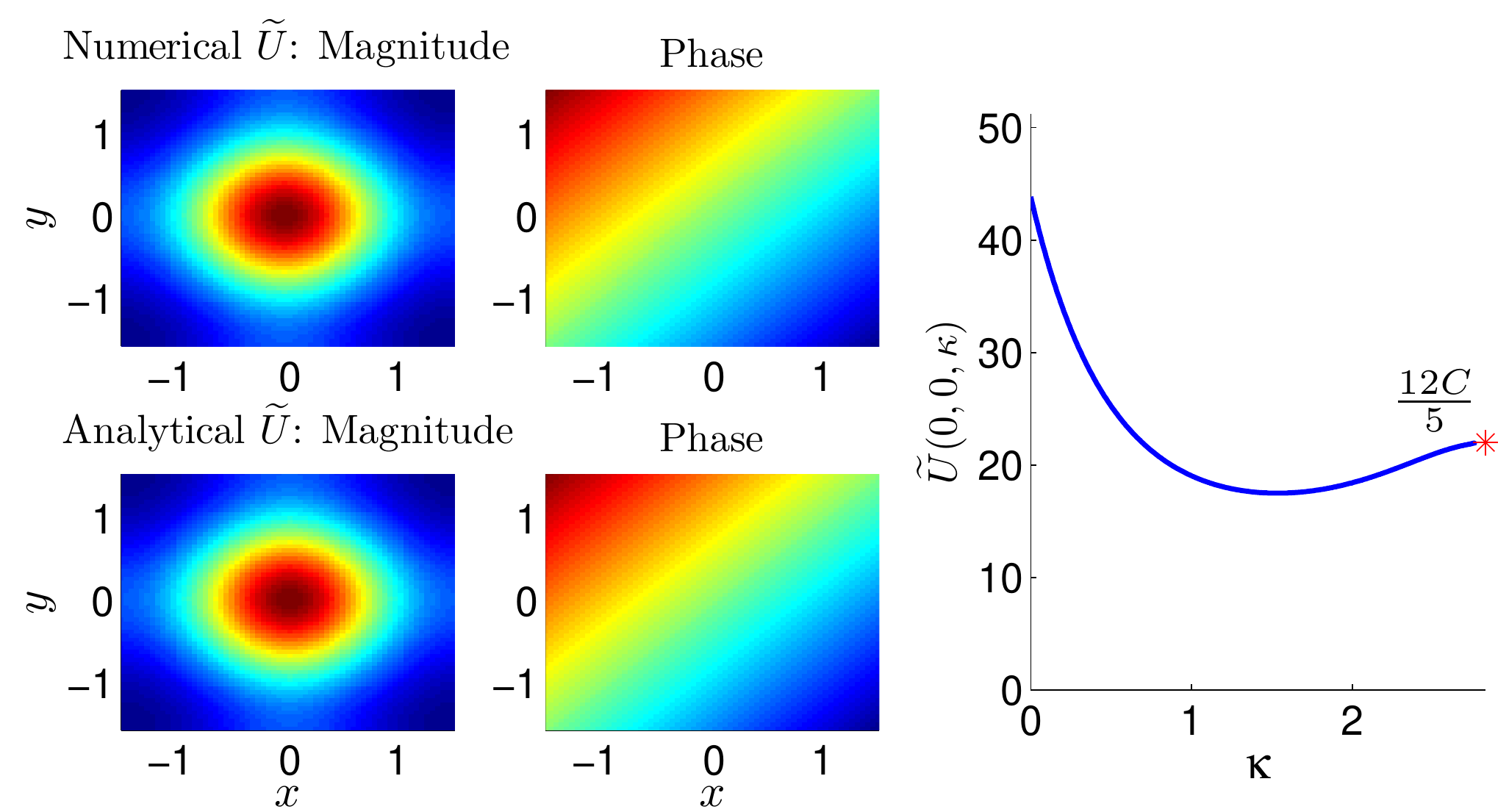}
\end{center}
\caption{(Color online) Numerical solutions of the outer integral
equation (\ref{Eq: Uorder1}) for the lattice (\ref{Eq: ExLat}) with
$\sigma=1$ and $\tan\theta = 1$. Left panels: solutions at the
singularity $\kappa=\kappa_0$; upper row: the numerical solution,
lower row: the analytical solution. Right panel: the solution versus
$\kappa$ at $x=y=0$. } \label{Fig: Numerics}
\end{figure}

Numerically we have confirmed the above outer-solution behavior near
the poles. As an example, the numerical results for the lattice
(\ref{Eq: ExLat}) with $\sigma=1$, $\tan \theta = 1$ and
$\mu_0=4.1264$ (edge of the semi-infinite gap) are shown in figure
\ref{Fig: Numerics}. For this line slope, $\kappa_0=2\sqrt{2}$. On
the top left the contour plots of $\widetilde{U}(x,y,\kappa_0)$ are
displayed. The corresponding analytic formula \eqref{Eq:
UtildeAsymptotics} with real $C$ is also shown at the bottom for
comparison. One can see that the two solutions match very well. On
the right side of Fig. \ref{Fig: Numerics}, the solution
$\widetilde{U}(0,0,\kappa)$ is shown. When $\kappa\to \kappa_0$,
$\widetilde{U}(0,0,\kappa)\to 22.05$, thus the fourth-order pole at
$\kappa=\kappa_0$ is numerically confirmed. If $\tan \theta = 2$,
then we find that $\widetilde{U}(0,0,\kappa)\to 4.04\times 10^4$ as
$\kappa\to \kappa_0=2\sqrt{5}$. Recalling the analytical formula
(\ref{Eq: UtildeAsymptotics}) and Bloch-mode normalization (which
boils down to $b(0,0)=1$ here), the constant $C$ is then inferred as
\begin{align}
C &= 9.19, \hspace{1.7cm} {\rm for}~ \tan\theta = 1,   \label{Catan1} \\
C &= 1.68 \times 10^4, \hspace{0.75cm} {\rm for}~ \tan\theta = 2.
\label{Catan2}
\end{align}
In section \ref{Sec: Eigenvalues} we are able to verify these values
of $C$ quantitatively, by comparing our predictions to numerical
calculations of a certain linear-stability eigenvalue.

The approach taken here of directly solving the outer integral
equation (\ref{Eq: Uorder1}) by numerical methods not only overcomes
the inadequacy of the recurrence relation (for two-dimensional
problems), but also has a number of additional advantages.
Specifically, compared with the computation of recurrence relations
in earlier works \cite{Hwang2011,Hwang2012, Akylas1997}, the direct
numerical solution of the outer integral equation, as was explained
above, is actually easier. In addition, this direct computation
gives directly the pole strength (\ref{Eq: UAsymptotics}) in the
outer solution, while in the previous approach the pole strength was
inferred indirectly from the recurrence solution. In view of these
advantages, it is concluded that this direct solving of the outer
integral equation also simplifies the exponential asymptotics
procedure.

\section{Solution near the poles}
\label{Sec: NearPole}

Based on experience from prior work \cite{Hwang2011,Hwang2012,
Akylas1997}, the actual poles in the solution $U(x,y,\kappa;
\epsilon)$ are expected to be $O(\epsilon)$ away from the real axis.
To determine the residues of these true poles, we need to analyze
the solution $U(x,y,\kappa; \epsilon)$ near these poles. In this
`inner' region, the reduced outer equation (\ref{Eq: Uorder1}) does
not hold, and one needs to work with the full equation (\ref{Eq:
UFull}) instead. For line solitons, it turns out that the analysis
of the inner solution nearly replicates that in earlier works
\cite{Hwang2011,Hwang2012,Akylas1997}, thus we shall keep this
discussion brief.

Focusing on the behavior of the solution $U(x,y,\kappa; \epsilon)$
in the inner region near the closest positive singularity,
\begin{equation}
\kappa_0 = 2\sqrt{p^2+q^2},
\end{equation}
we introduce an inner variable, $\xi = (\kappa -
\kappa_0)/\epsilon$, that is $\kappa = \kappa_0 + \epsilon \xi$,
with $\xi =O(1)$. In this region we expand the solution to integral
equation \eqref{Eq: UFull} as
\begin{equation}
U(x,y,\kappa; \epsilon) = \frac{\re^{-\ri \kappa_0
w}}{\epsilon^4}\left\{ \Phi_0(\xi) b(x,y) + \ri \epsilon \xi
\Phi_0(\xi) \nu(x,y) + O(\epsilon^2)\right\}. \label{Eq: PoleExpac}
\end{equation}
Here the order of the solution near the pole $U = O(\epsilon^{-4})$
is chosen to match the large-$\xi$ behavior of the solution
$\Phi_0(\xi) = O(\xi^{-4})$ (see Eq. (\ref{Eq: Phi0largexi}) below).

The dominant contribution from the double integral in equation
\eqref{Eq: UFull} comes from the three regions: (i) $r \approx 0$,
$s\approx 0$, (ii) $r \approx \kappa$, $s\approx 0$, and (iii) $r
\approx \kappa$, $s\approx \kappa$. In the first region,

\begin{align}
& U(x,y, r - s ) \approx U(x,y,s) \approx \frac{\alpha \beta}{2} b(x,y), \\
& U(x,y, \kappa - r) \approx \frac{\re^{-\ri \kappa_0 w}}{\epsilon^4} \Phi_0\left(\frac{\kappa -\kappa_0 - r}{\epsilon}\right) b(x,y), \\
& \cosh\frac{\pi\beta \kappa}{2\epsilon}\left/ \cosh\frac{\pi\beta
(\kappa-r)}{2\epsilon} \right.   \approx \re^{\pi \beta r /2
\epsilon}.
\end{align}
Changing variables $\tilde{s} = s/\epsilon$, $\tilde{r} = r/\epsilon$ and using the fact that
\begin{equation}
\AllInt \sech(x-y) \sech(y) \hspace{0.05cm} \rd y = 2x
\hspace{0.05cm} \csch(x),
\end{equation}
the contribution from the double integral in the first region can be
readily obtained. Contributions from the other two regions can be
calculated in a similar manner, and they turn out to be the same as
that from the first region.

With the change of variables $U(x,y,\xi) = \re^{-\ri \kappa_0
w}\widehat{U}(x,y,\xi)$, we then find that near the singularity the
full integral equation \eqref{Eq: UFull} reduces to the inner
equation
\begin{align}
L_0 \widehat{U} + \epsilon \xi L_1\widehat{U} & + \epsilon^2 (\eta - \xi^2) \widehat{U}
+\frac{3}{2\epsilon^2} \sigma\alpha^2\beta^2 b(x,y)^3 \notag \\
&\times \AllInt \omega \re^{\pi\beta\omega/2}
\csch\left(\frac{\pi\beta\omega}{2} \right) \Phi_0(\xi - \omega) \rd
\omega = 0. \label{Eq: InnerEquation}
\end{align}
After substituting in the expansion \eqref{Eq: PoleExpac}, at
$O(\epsilon^{-4})$ and $O(\epsilon^{-3})$ the above equation is
automatically satisfied; and at $O(\epsilon^{-2})$ the equation
governing $\Phi_0(\xi)$ is found from the solvability condition as
\begin{align}
(1+ \beta^2 \xi^2)\Phi_0 - 3\beta^2  \AllInt \omega
\re^{\pi\beta\omega/2} \csch\left(\frac{\pi\beta\omega}{2} \right)
\Phi_0(\xi - \omega) \rd \omega =0.  \label{Eq: phi0}
\end{align}
This is a linear homogeneous Fredholm integral equation which has
been solved before \cite{Hwang2011,Akylas1997}. Its analytic
solution in the region $\left|\Imag(\xi)\right| \geq 1/\beta$ is
\begin{equation}
\Phi_0(\xi) = \frac{6\beta^4}{1+ \beta^2 \xi^2} \int_0^{\pm \ri
\infty} \frac{1}{\sin^2 s}\phi(s) \re^{-s\beta \xi} \rd s,
\label{Eq: Phi0solution}
\end{equation}
where the plus sign in the contour is for $\Imag(\xi)\leq -1/\beta$,
the minus sign for $\Imag(\xi)\geq 1/\beta$,
\begin{equation}
\phi(s) = C\left( \frac{2}{\sin s} + \frac{\cos^2 s}{\sin s} -
\frac{3s \cos s}{\sin^2s}\right),
\end{equation}
and $C$ is a complex constant. Clearly this solution has simple-pole
singularities at $\xi = \pm \ri /\beta$. Since the integral of
(\ref{Eq: Phi0solution}) at these points is equal to $-C/6$, we see
that
\begin{equation}
\Phi_0(\xi) \sim -\frac{C\beta^4}{1+ \beta^2 \xi^2},~~{\rm for} ~\xi
\to  \pm \frac{\ri}{\beta},
\end{equation}
thus the pole has strength $\pm \ri \beta^3 C/2$ at $\xi =\pm
\ri/\beta$. After substituting this back into equation \eqref{Eq:
PoleExpac} and changing variables to $K =\kappa_0/\epsilon+ \xi $,
we find that $U(x,y,K)$ has simple poles at $K= \kappa_0/\epsilon
\pm i/\beta$, and
\begin{equation}
U(x,y,K) \sim \pm  \frac{\ri \beta^3C}{2\epsilon^4} \re^{-\ri
\kappa_0 w} b(x,y) \frac{1}{K - \left(\frac{\kappa_0}{\epsilon} \pm
\frac{\ri}{\beta}\right)}   ~~~{\rm for} ~K \rightarrow
\frac{\kappa_0}{\epsilon} \pm \frac{\ri}{\beta}. \label{Eq: Pole}
\end{equation}
Finally, from Eq. (\ref{Eq: Psihat}), we obtain the local behavior
of $\widehat{\psi}(x,y,K)$ near the simple poles $K
=\kappa_0/\epsilon\pm i/\beta$ as
\begin{equation}
\widehat\psi(x,y,K) \sim \frac{\beta^3 C}{\epsilon^3}
e^{-\pi\beta\kappa_0/2\epsilon}e^{\pm W_0/\beta}
\frac{e^{-i\kappa_0(w+w_0)}}{K-\left(\frac{\kappa_0}{\epsilon} \pm
\frac{i}{\beta}\right) }b(x,y), \quad  K\to \frac{\kappa_0}
{\epsilon} \pm \frac{i}{\beta},
\end{equation}
where
\[w_0=W_0/\epsilon.\]
Moreover, from the symmetry of the Fourier transform
\begin{equation}
\widehat\psi(x,y, K)={\widehat\psi}^*(x,y, -K^*)
\end{equation}
for real functions $\psi(x,y,W)$, we also deduce the local behavior
of $\widehat{\psi}(x,y,K)$ near the simple poles
$K=-\kappa_0/\epsilon\pm i/\beta$.

From the above local analysis, the residues of these poles are only
determined up to a constant multiple ($C$ is unknown at the moment).
To determine $C$, we match the large-$\xi$ asymptotics of the above
inner solution for $U$ with the outer solution for $U$ away from the
singularities, in the matching region $1\ll |\xi|\ll \epsilon^{-1}$.
To this end, we note that for $|\xi|\gg 1$ the main contribution of
the integral in solution \eqref{Eq: Phi0solution} comes from the
region $s\approx 0$ where $\phi(s) \sim \frac{2}{5} C s^3$. This
yields
\begin{equation} \label{Eq: Phi0largexi}
\Phi_0(\xi) \sim \frac{12 C}{5} \frac{1}{\xi^4},~~{\rm for} ~ |\xi|
\gg 1.
\end{equation}
Putting this back into equation \eqref{Eq: PoleExpac}, we find the
following large-$\xi$ behavior for the inner solution,
\begin{equation}
U(x,y,\kappa) \sim \frac{12 C}{5} \frac{b(x,y)}{(\kappa -
\kappa_0)^4}\re^{-\ri \kappa_0 w}. \label{Eq: InnerAsymptotics}
\end{equation}
As discussed in section \ref{Sec: Near0}, this behavior matches to
the outer solution with $|\kappa -\kappa_0|\ll~1$.

\section{Inversion of Fourier transform and true line solitons}
\label{Sec: Inversion}

We now take the inverse Fourier transform of (\ref{Eq: FT}),
\begin{equation}  \label{Eq: IFT}
\psi(x,y,W)=\int_{\cal C} \widehat{\psi}(x,y,K) e^{iKW} dK
\end{equation}
in order to determine the tail behaviors in the physical solution
$\psi(x,y,W)$. Here $\widehat{\psi}(x,y,K)$ is given by Eq.
(\ref{Eq: Psihat}). As explained in \cite{Hwang2011}, if we require
this physical solution to decay upstream ($w\to -\infty$), then the
contour $\cal C$ in this inverse Fourier transform should be taken
along the line $\Imag(K)=-1/\beta$ and pass below the poles $K= \pm
\kappa_0/\epsilon - i/\beta$. It should also pass above the pole
$K=-i/\beta$ of the $\sech(\pi\beta K/2)$ term in Eq. (\ref{Eq:
Psihat}). Then when $w\gg 1$ (downstream), by completing the contour
$\cal C$ with a large semicircle in the upper half plane, we pick up
dominant contributions from the pole singularities at
$K=\pm\kappa_0/\epsilon -i/\beta$ and $K=i/\beta$. Collecting these
pole contributions, the wave profile of the solution far downstream
is then found to be
\begin{align}
\psi\sim & 2\epsilon \alpha \hspace{0.05cm} e^{-(W-W_0)/\beta}b(x,y)
\nonumber
\\&+\frac{4\pi\beta^3 C_0}{\epsilon^3}
e^{- \pi \beta \kappa_0 / 2\epsilon}\sin(\kappa_0 w_0-\Theta_0)
e^{(W-W_0)/\beta}b(x,y), \quad \,\, w \gg 1/\epsilon, \label{Eq:
Downstream}
\end{align}
where $C_0>0$ and $\Theta_0$ are the amplitude and phase of the
constant $C$ (which is complex in general), i.e., $C = C_0 \re^{\ri
\Theta_0}$.

For this solution to be a line soliton, the growing term in
(\ref{Eq: Downstream}) must vanish so $\sin \left( \kappa_0 w_0 -
\Theta_0 \right) = 0$. Thus, there are two allowable locations for
line solitons (relative to the lattice),
\begin{equation}  \label{Eq: w0}
w_0 = \Theta_0 / \kappa_0, \quad (\pi + \Theta_0)/\kappa_0.
\end{equation}
Line solitons at these two locations are called onsite and offsite
solitons respectively. For the particular lattice (\ref{Eq: ExLat})
and $\sigma=1$, these two line solitons near the edge of the
semi-infinite gap with $\tan\theta=1, 2$ are displayed in Fig.
\ref{Fig:LSolitons}. For these solitons, $\Theta_0=0$ since $C$ is
real positive (see equations (\ref{Catan1}) and (\ref{Catan2})).

The above results show that for any rational slope $\tan\theta$, two
line solitons with envelope locations (\ref{Eq: w0}) exist in a
general two-dimensional lattice. What if the slope is irrational?
Treating an irrational number as the limit of a rational number
$p/q$ with $p, q \to \infty$, then $\kappa_0=\sqrt{p^2+q^2} \to
\infty$, hence the growing tail downstream in (\ref{Eq: Downstream})
vanishes. This suggests that for an irrational slope, line solitons
exist for arbitrary envelope positions $w_0$. But we cannot
numerically verify this conjecture since irrational numbers cannot
be represented accurately on computers.

Finally, if one wishes to obtain line wave packets $\psi(x,y,W)$
which decay for $w\to +\infty$ but contain a growing tail for $w\ll
-1$, then the contour $\cal C$ in the inverse Fourier transform
should be taken along the line $\Imag(K)=1/\beta$ and pass above the
poles $K= \pm \kappa_0/\epsilon + i/\beta$ and below the pole
$K=i/\beta$. Then when $w\ll -1$, by completing the contour $\cal C$
with a large semicircle in the lower half plane and picking up
dominant pole contributions, the wave profile of the solution is
found to be
\begin{align}
\psi\sim & 2\epsilon \hspace{0.04cm} \alpha \hspace{0.05cm}
e^{(W-W_0)/\beta}b(x,y)
\nonumber \\
& -\frac{4\pi\beta^3 C_0}{\epsilon^3} e^{- \pi \beta \kappa_0 /
2\epsilon}\sin(\kappa_0 w_0-\Theta_0) e^{-(W-W_0)/\beta}b(x,y),
\quad \,\, w \ll -1/\epsilon. \label{Eq: Downstream2}
\end{align}
This `flipped' wave solution will be useful when we construct
multi-line-soliton bound states in the next section.

\section{Construction of multi-line-soliton bound states}
\label{Sec: Boundstates}

The asymptotic tail formula (\ref{Eq: Downstream}) can be used not
only to determine the locations of line solitons, but also to
construct multi-line-soliton bound states. These bound states are
analytically constructed by matching the downstream growing tail of
a line wavepacket with the upstream decaying tail of another line
wavepacket. This technique has been used for the construction of
one-dimensional multi-soliton bound states before
\cite{Akylas2012,Hwang2012,Akylas1997}.  Here we apply the same
principle to the construction of multi-line soliton solutions in a
general two-dimensional lattice potential.

Consider the asymptotic expansion of two line wavepackets centered
at $w_1= W_1/\epsilon$ (left) and $w_2= W_2/\epsilon $ (right)
receptively. The left wavepacket decays for $w - w_1 \ll -1$ and has
a growing exponential tail for $w-w_1 \gg 1$, and the right
wavepacket decays for $w-w_2 \gg 1$ and has a growing exponential
tail for $w - w_2 \ll -1$. For line-soliton bound states, the
decaying and growing tails of the two wavepackets must match in the
region $w_1 \ll w \ll w_2$. In this matching region, the left
wavepacket's asymptotics is given by (\ref{Eq: Downstream}) with
$w_0$ replaced by $w_1$, and the right wavepacket's asymptotics is
given by (\ref{Eq: Downstream2}) with $w_0$ replaced by $w_2$.
Matching of these asymptotics results in the following system of
equations
\begin{align*}
2\epsilon \alpha \re^{-(W-W_1)/\beta} b(x,y) &= \mp \frac{4 \pi
\beta^3C_0}{\epsilon^3}\re^{ - \pi \beta \kappa_0 / 2\epsilon} \sin
\left( \kappa_0 w_2 - \Theta_0 \right) \re^{-(W-W_2)/\beta} b(x,y),
\\
2\epsilon \alpha \re^{(W-W_2)/\beta} b(x,y)&= \pm \frac{4 \pi
\beta^3C_0} {\epsilon^3} \re^{ - \pi \beta \kappa_0 / 2\epsilon}
\sin \left( \kappa_0 w_1 - \Theta_0 \right) \re^{(W-W_1)/\beta}
b(x,y),
\end{align*}
where the $\mp$ comes from the possible $\pi$ phase shift between
the two wavepackets. After simplification, these matching conditions
read
\begin{equation}
\sin\left( \kappa_0 w_1 - \Theta_0 \right) = -\sin\left( \kappa_0
w_2 - \Theta_0 \right) = \pm \frac{\alpha\epsilon^4}{2 \pi \beta^3
C_0 } \re^{\pi\beta\kappa_0/2\epsilon} \re^{\epsilon
(w_1-w_2)/\beta}. \label{Eq: Centers}
\end{equation}
With a change of variables $\widehat{w}_1=-(w_1-\Theta_0/\kappa_0)$,
$\widehat{w}_2=w_2-\Theta_0/\kappa_0$, the above matching conditions
then become almost identical to the ones derived in
\cite{Akylas2012} before. As has been explained there, this system
of equations admits an infinite number of solutions for each fixed
$\epsilon>0$. Varying $\epsilon$, then infinite families of two-line
soliton bound states are obtained.

Note that for equations \eqref{Eq: Centers} to have a solution, the
right-hand side must have magnitude less than one. Since this is not
the case as $\epsilon \rightarrow 0$ for any finite distance
$w_2-w_1$ between the two line wavepackets, bifurcations of these
bound states occur at finite amplitude away from the band edge. In
figure \ref{Fig: MultiSoliton}, we show a particular family of bound
states in the lattice (\ref{Eq: ExLat}) with $\sigma=1$ and $\tan
\theta = 1$. This family bifurcates at $\mu \approx 4.073$ (near the
edge $\mu_0=4.1264$ of the semi-infinite gap). As predicted by the
analysis of equation (\ref{Eq: Centers}) in \cite{Akylas2012}, this
solution family contains three connected branches, and their power
curve is shown in Fig. \ref{Fig: MultiSoliton} (upper left panel).
Here the power is calculated over one period along the line-soliton
direction $w'= x\cos\theta + y\sin\theta$, which is $\pi \sqrt{p^2 +
q^2}$ in the present case. Profiles of bound states at $\mu=4.0639$
of the three power branches are displayed in Fig. \ref{Fig:
MultiSoliton} (A,B,C). The bound state in the A panel comprises
roughly two onsite line solitons, the one in the B panel comprises
roughly an onsite and an offsite line solitons, and the bound state
in the C panel comprises roughly two offsite line solitons.

\begin{figure}[htbp!]
\begin{center}
\includegraphics[width=5in]{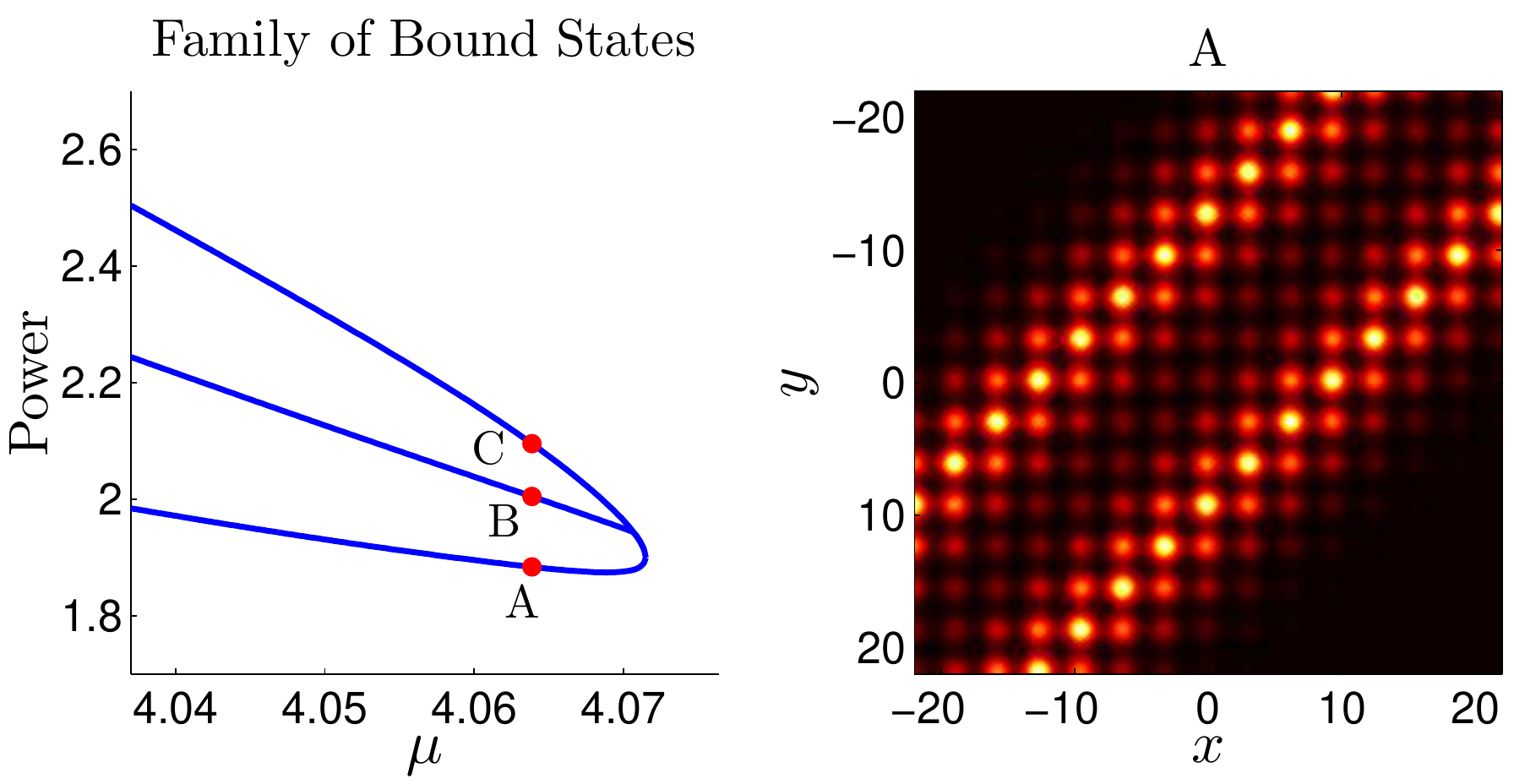}
\includegraphics[width=5in]{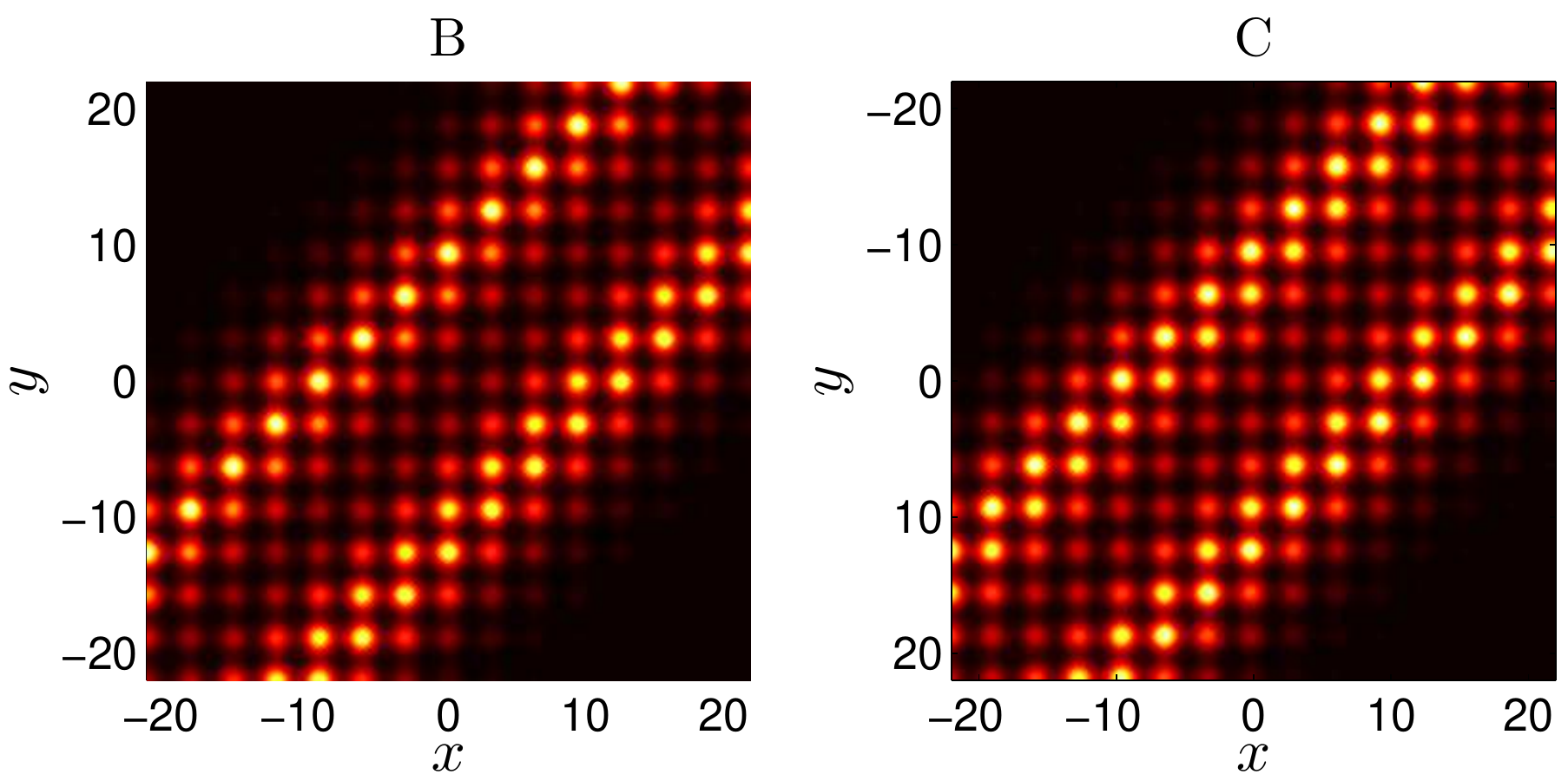}
\end{center}
\caption{(Color online) Power curve and solution profiles in a
family of two-line soliton bound states in the lattice (\ref{Eq:
ExLat}) with $\sigma=1$ and $\tan \theta = 1$.} \label{Fig:
MultiSoliton}
\end{figure}

We would like to point out that the above construction of infinite
families of line-soliton bound states was performed for general
two-dimensional lattices. This contrasts the earlier work in
\cite{Akylas2012,Hwang2012} where such construction was made only
for symmetric one-dimensional lattices (where $\Theta_0=0$). From
the above calculations, it is now clear that the previous derivation
of multi-soliton bound states in \cite{Akylas2012,Hwang2012} can be
extended to general one-dimensional lattices, too.

\section{Connection to zero-eigenvalue bifurcations}
\label{Sec: Eigenvalues}

In this section, we consider the linear stability eigenvalues of
single-line solitons as obtained in section \ref{Sec: Inversion}.
Our interest lies in the pair of exponentially small eigenvalues of
these solitons, which bifurcate out from the origin at the band
edge; the associated eigenfunctions have zero wavenumber along the
line-soliton direction (i.e., are periodic along the line direction
with the period matching that of the line soliton). Such eigenvalues
can be analytically calculated from the tail asymptotics (\ref{Eq:
Downstream}) of line wavepackets which we have derived. Thus, by
comparing the analytical formula of these eigenvalues with the
numerically computed eigenvalues, we can quantitatively verify the
formula in Eq. (\ref{Eq: Downstream}) for the exponentially small
growing tails. We do caution, however, that we are ignoring other
eigenvalues whose eigenfunctions have nonzero wavenumber along the
line-soliton direction, and those eigenvalues are often more
unstable \cite{Yang2011,PeliYang2012}. Thus the eigenvalue
calculation in this section does not constitute a full stability
analysis.

The present calculation of zero-eigenvalue bifurcation closely
parallels that in one-dimensional problems
\cite{Akylas2000,Hwang2011,Hwang2012}. Let $\psi_s(x,y)= \psi(x,y;
w_{0s})$ be a single-line soliton solution of equation \eqref{Eq:
psi} with center at $w_0 = w_{0s}$, which decays to zero as $w
\rightarrow \pm \infty$ and is periodic along the line direction,
$w'= x\cos\theta + y\sin\theta$, with period matching that of the
Bloch wave. Perturbing this line soliton by normal modes
\begin{equation}
\Psi = \re^{-\ri \mu t} \left[ \psi_s + (v+\varphi)\re^{\lambda t}
+(v^* - \varphi^*)\re^{\lambda^*t} \right],
\end{equation}
with $v,~\varphi \ll 1$, we obtain the linear-stability eigenvalue
problem
\begin{equation} \label{e:eigenvalproblem}
{\cal L}_0{\cal L}_1v=-\lambda^2v,
\end{equation}
where
\[
{\cal L}_0 =\nabla^2 +\mu  - V(x,y)  + \sigma \psi_s^2, \quad  {\cal
L}_1 =\nabla^2 +\mu  - V(x,y)  + 3\sigma \psi_s^2,
\]
and $\lambda$ is the stability eigenvalue. Since the bifurcated
eigenvalue $\lambda$ is small near band edges, we expand the
eigenfunction $v$ into a perturbation series
\begin{equation}  \label{vexpansion}
v=v_0+\lambda^2v_1+\lambda^4v_2+\dots.
\end{equation}
Inserting this expansion into Eq. (\ref{e:eigenvalproblem}), at
$O(1)$ we get
\begin{equation}  \label{L0L1v0}
L_0L_1v_0=0.
\end{equation}
A solution to this equation is
\begin{equation} \label{Eq: v0}
v_0=\left(\partial\psi/\partial w_0\right)_{w_0=w_{0s}}.
\end{equation}
Recalling the perturbation series expansion of $\psi(x,y; w_0)$ in
Eq. (\ref{Eq: Aseries}) as well as the envelope solution in Eq.
(\ref{Eq: A}), we find that
\[
v_0 \sim -\frac{\epsilon^2\alpha}{\beta}
\sech\frac{\epsilon(w-w_{0s})}{\beta} \tanh
\frac{\epsilon(w-w_{0s})}{\beta} b(x,y).
\]
From this equation it is seen that the eigenfunction $v$ is periodic
along the line-soliton direction $w'= x\cos\theta + y\sin\theta$,
with period matching that of the Bloch wave (and the line soliton),
so the net wavenumber of this eigenfunction along the line-soliton
direction is zero.

From the large-$w$ asymptotics \eqref{Eq: Downstream} of the
solution $\psi(x,y; w_0)$, we see that when $w \gg 1/\epsilon$,
$v_0$ contains a growing tail,
\begin{equation}
v_0  \sim \pm \epsilon^{-3} 4 \pi \beta^3 \kappa_0 C_0
\re^{-\pi\beta \kappa_0 / 2 \epsilon}
\re^{\epsilon(w-w_{0s})/\beta}b(x,y). \label{Eq: EigTail0}
\end{equation}
Here the plus and minus sign correspond to the onsite and offsite
line solitons, respectively. This growing tail must be balanced by
the higher-order terms in the expansion (\ref{vexpansion}), thereby
yielding an analytical formula for the eigenvalue $\lambda$.
Carrying out this program as in
\cite{Hwang2011,Hwang2012,Akylas2000}, we find that
\begin{equation}  \label{Eq: lambda}
\lambda^2 =  \mp C_0 \frac{32 \pi \kappa_0 \beta^4 }{\alpha
\epsilon} \re^{-\pi \beta \kappa_0 / 2 \epsilon}.
\end{equation}
Thus this eigenvalue is stable for onsite line solitons and unstable
for offsite ones, and its magnitude is exponentially small.

\begin{figure}[htbp!]
\begin{center}
\includegraphics[width=4.5in, height=2in]{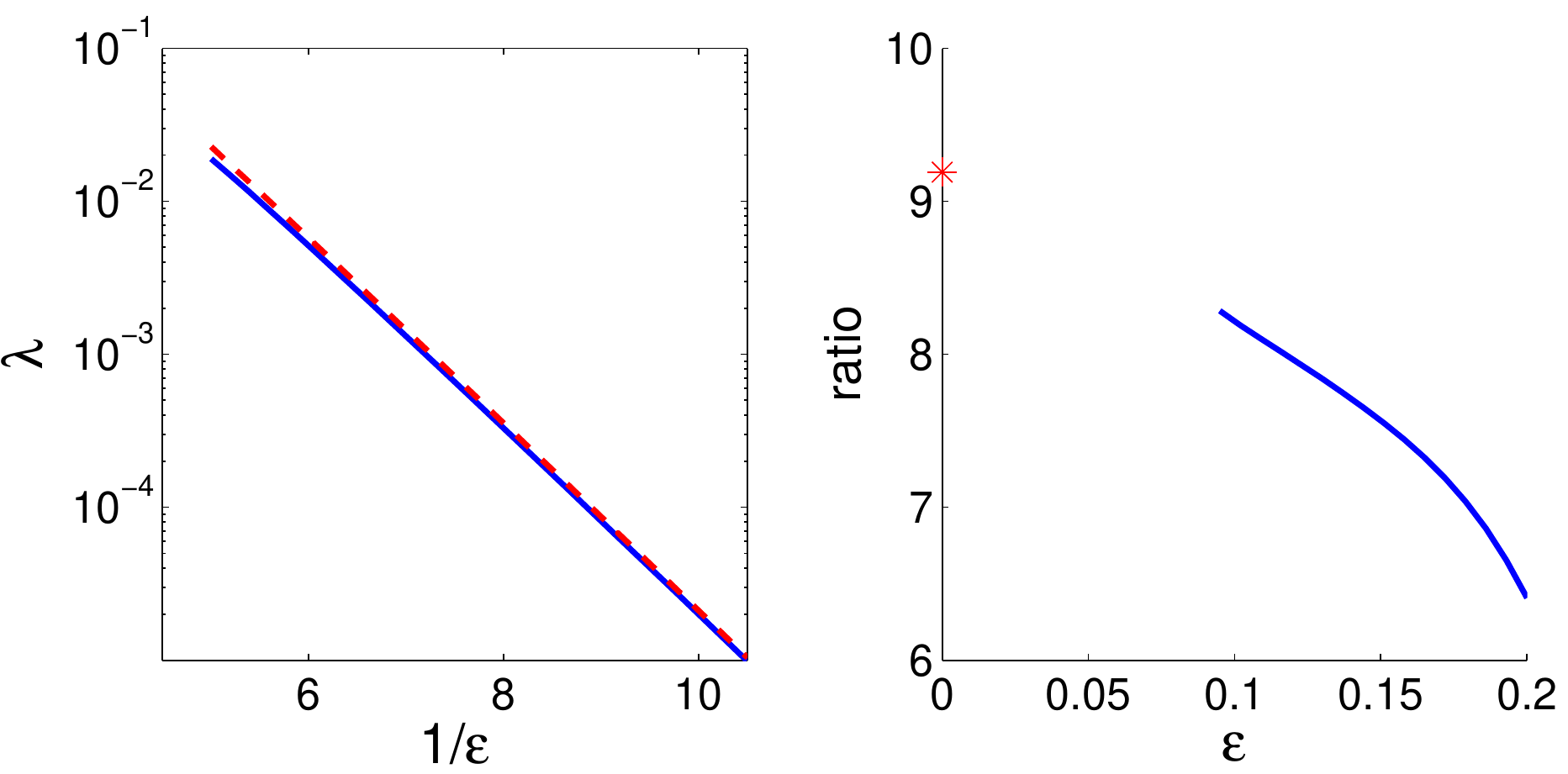}

\includegraphics[width=4.5in, height=2in]{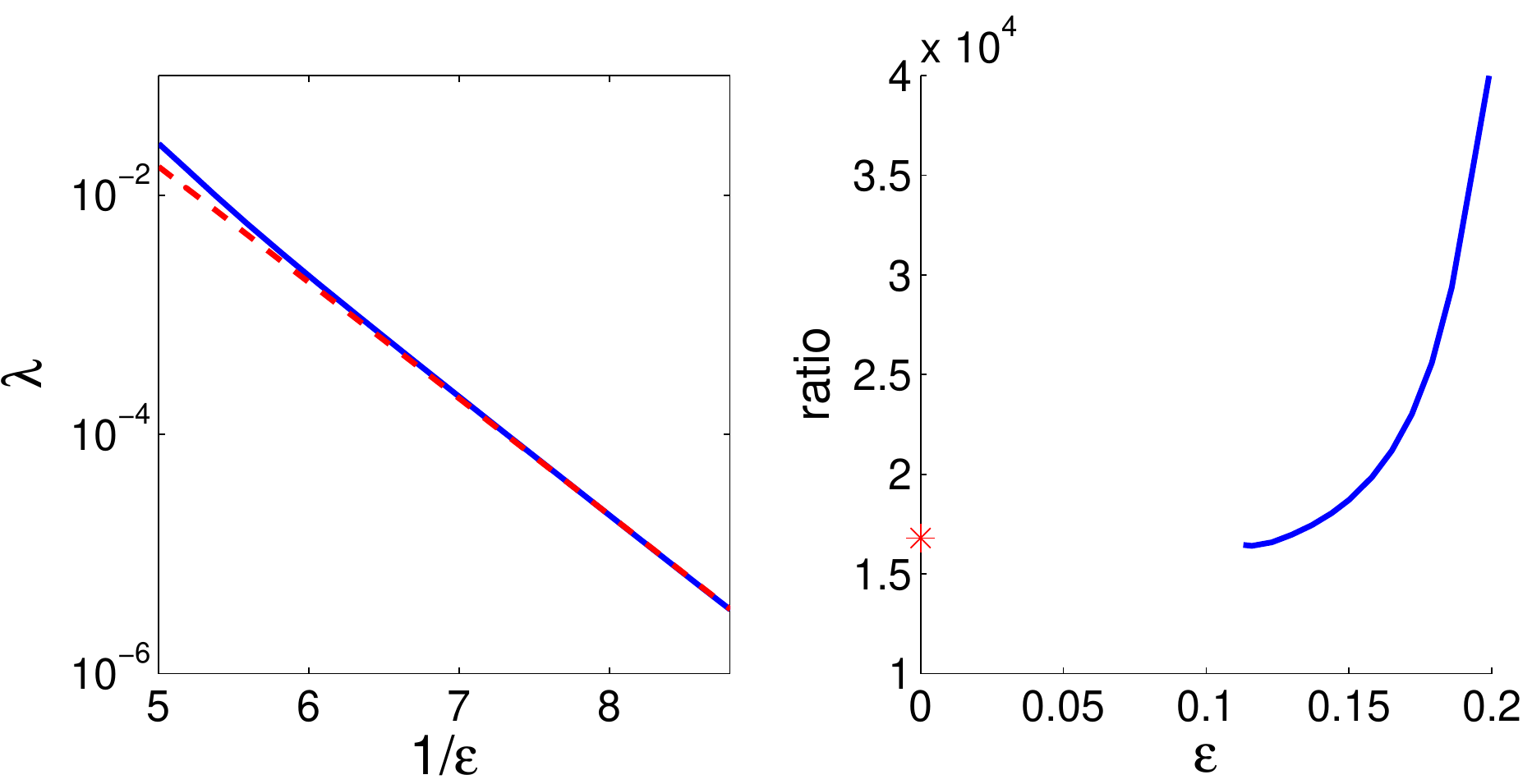}
\caption{Eigenvalue comparison for offsite line solitons with
$\sigma=1$ and lattice (\ref{Eq: ExLat}): (top) $\tan\theta = 1$;
(bottom) $\tan\theta = 2$. (Left) Numerically calculated $\lambda$
value in solid blue and the asymptotic approximation in dashed red.
(right) The value for C left from taking the ratio of the numerical
eigenvalues and the asymptotic expression as well as the predicted
limiting value (red star) from equations (\ref{Catan1}) and
(\ref{Catan2}) in section \ref{Sec: Near0}.} \label{Fig: LambdaTan1}
\end{center}
\end{figure}

A comparison of the analytical prediction (\ref{Eq: lambda}) for the
eigenvalue to numerical results for offsite line solitons is
presented in Fig. \ref{Fig: LambdaTan1} for the lattice (\ref{Eq:
ExLat}) with $\sigma=1$ and $\tan\theta = 1, 2$. It is seen that the
analytical and numerical eigenvalues agree with each other very
well. Notice that the analytical eigenvalue formula contains the
constant $C_0$, and the ratio of $\lambda^2/[32 \pi\kappa_0 \beta^4
\re^{-\pi \beta \kappa_0 / 2 \epsilon}/\alpha \epsilon]$ approaches
$C_0$ when $\epsilon\to 0$. We have plotted this ratio for
numerically obtained eigenvalues in the right panels of Fig.
\ref{Fig: LambdaTan1}. It is seen that as $\epsilon\to 0$, this
ratio indeed approaches the $C$ value obtained from equations
(\ref{Catan1}) and (\ref{Catan2}) in section \ref{Sec: Near0}. Thus
our formula (\ref{Eq: Downstream}) for the exponentially-small
growing tails in line wavepackets is fully verified quantitatively.

\section{Further examples with asymmetric potentials}

\begin{figure}[htbp!]
\begin{center}
\includegraphics[width=5in]{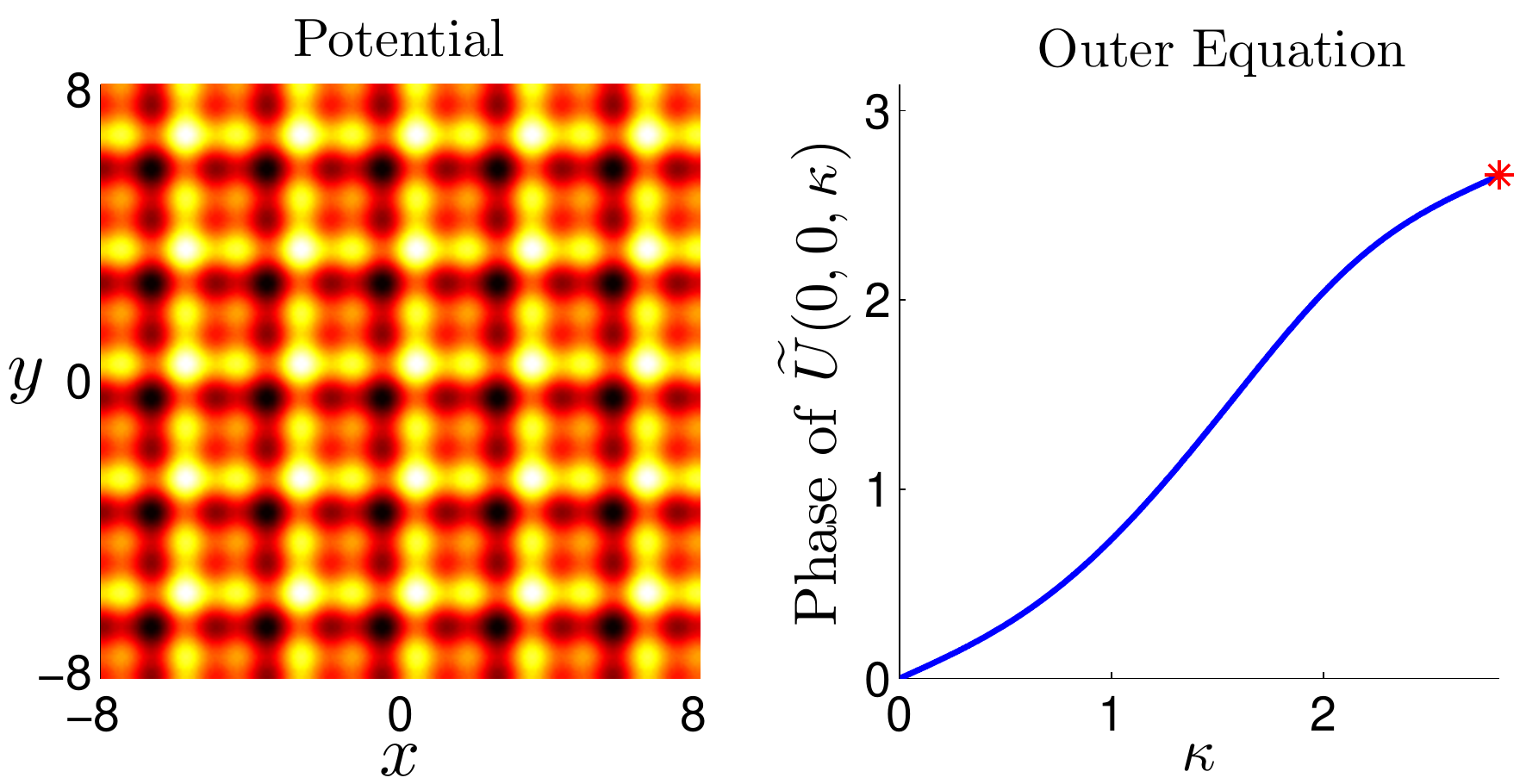}

\vspace{0.04cm}
\includegraphics[width=5in]{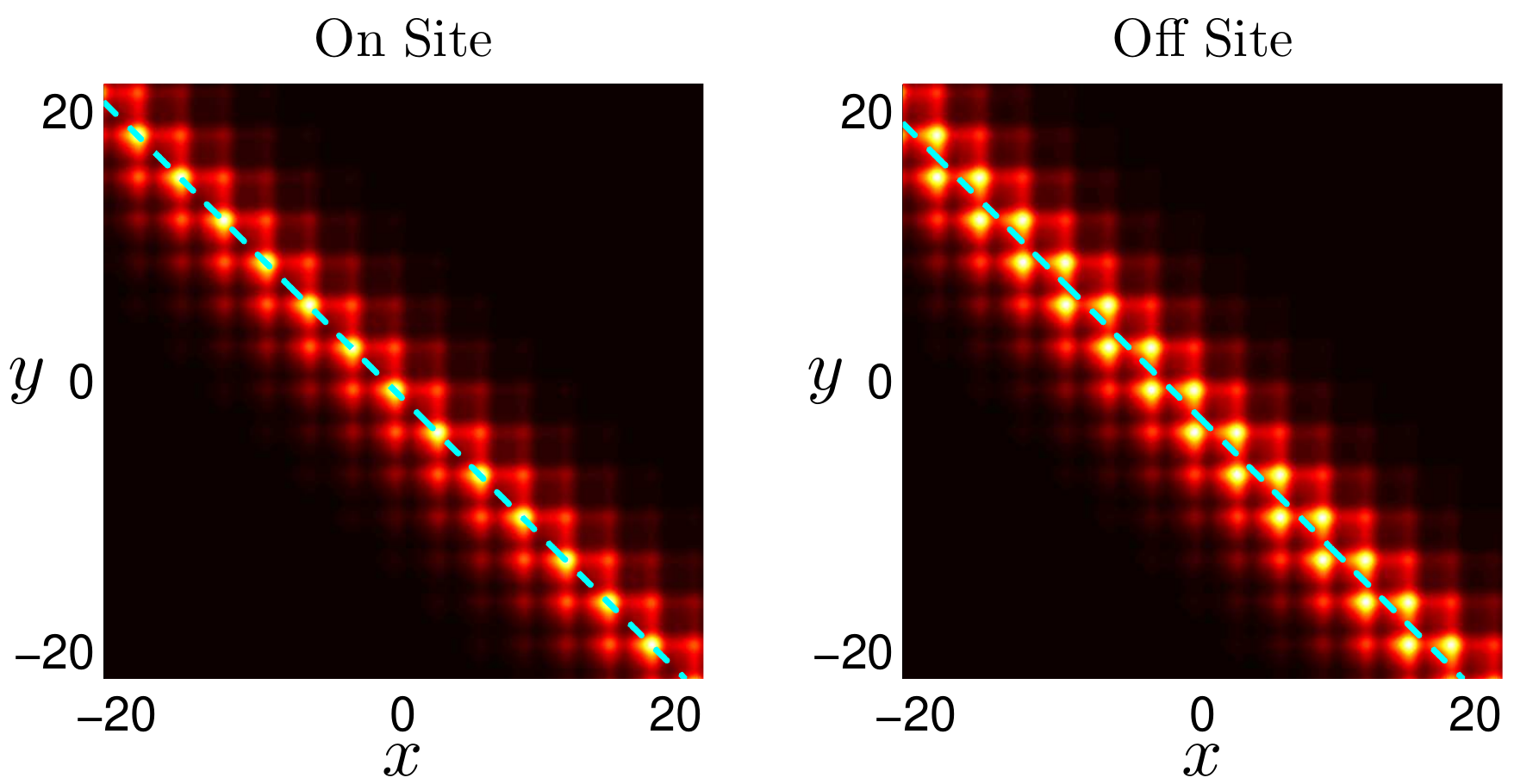}
\end{center}
\caption{(Color online) Line solitons in the asymmetric lattice
(\ref{Eq: asymlat}) with $\sigma=1$ and $\tan\theta=-1$. (Top left)
the potential (\ref{Eq: asymlat}). (Top right) the phase calculated
from the ``outer" integral equation \eqref{Eq: Uorder1}; the
asterisk marks the phase value $\Theta_0$ at the singularity
$\kappa_0=2\sqrt{2}$. (Bottom) onsite and offsite line solitons for
$\epsilon = 0.35$ as well as the predicted centers of the solitons
(dashed).} \label{Fig: AsymSolitons}
\end{figure}

To help demonstrate the generality of our theory, we now consider
the asymmetric lattice
\begin{equation}  \label{Eq: asymlat}
V(x,y) = \frac{3}{2}(\sin 2x+\sin 4x +\sin 2y+\sin 4y),
\end{equation}
which is shown in Fig. \ref{Fig: AsymSolitons} (top left panel). For
this lattice, the edge of the semi-infinite gap is $\mu_0 =
-0.6891$. With soliton inclination $\tan \theta = -1$ and
nonlinearity coefficient $\sigma=1$, we find the two soliton
families which bifurcate off this band edge (see Fig. \ref{Fig:
AsymSolitons}, bottom panel). In this case numerical solution of the
``outer" integral equation \eqref{Eq: Uorder1} yields a complex
value for $C = C_0\re^{\ri \Theta_0}$,
\begin{equation}
C_0 = 8.25, \quad \Theta_0 = 2.66,
\end{equation}
see Fig. \ref{Fig: AsymSolitons} (top right panel). This $\Theta_0$
value in turn gives the center of the solitons, from formulae
(\ref{Eq: w0}), as $w_0=0.94$ and $2.05$ for onsite and offsite,
respectively (up to periodic repetitions in the lattice). Comparison
of this prediction with the envelope locations of numerically
obtained solitons shows good agreement (see Fig. \ref{Fig:
AsymSolitons}, bottom panel).

\section{Line solitons bifurcated from interior points of Bloch
bands}  \label{Sec: Xpoint}

In previous sections, we studied line solitons that bifurcate from
edges of Bloch bands. But line solitons can also bifurcate from
high-symmetry points inside Bloch bands, as has been reported
numerically and experimentally before
\cite{Wang2007,Yang2011,PeliYang2012}. Here the high-symmetry points
are points on the dispersion surface $\mu=\mu(k_x,k_y)$ where
$\partial\mu/\partial k_x=\partial\mu/\partial k_y=0$, with $k_x,
k_y$ being the Bloch wavenumbers in the $x$ and $y$ directions. For
line solitons bifurcated from such interior points of Bloch bands,
however, resonance with Bloch modes may occur. Such resonance
excites Bloch-wave tails that are non-vanishing in the direction
perpendicular to the line and hence makes the line ``soliton"
nonlocal \cite{Boyd}. To obtain true line solitons inside Bloch
bands, this resonance must be absent, a requirement that poses
strong restrictions on the angles of line solitons. Indeed, we will
see below that inside Bloch bands, line solitons at only a couple of
special angles are admissible.

First we consider a concrete example with the lattice
\begin{equation}
V(x,y) = 6\sin^2 x + 4 \sin^2y.
\label{Eq: Xlattice}
\end{equation}
Inside the first Bloch band, $\mu \in [3.6080, \hspace{0.05cm}
4.1565]$, of this lattice there is an X-symmetry point
\begin{equation}  \label{Eq: Xpoint}
(k_{x0}, \ k_{y0}, \ \mu_0)=(0, \ 1, \ 3.9529),
\end{equation}
whose Bloch mode is $\pi$-periodic in $x$ and $2\pi$-periodic in
$y$. The dispersion surface near this X-point is saddle-shaped. For
line solitons bifurcating from this X-point,
$\mu=\mu_0+\eta\epsilon^2$, $\eta=\pm 1$, and $\epsilon \ll 1$.
Taking $\epsilon=0.2$, the level curves of the dispersion surface at
$\mu(k_x,k_y)=\mu_0+\eta\epsilon^2$ for $\eta=\pm 1$ are displayed
in Fig. \ref{Fig: XSoliton} (left panels, solid lines).

Suppose the angle of the line soliton with the $x$-axis is $\theta$.
Then at the $\mu$ value of the line soliton, linear Bloch modes that
are periodic along this line direction, with the period matching
that of the X-point Bloch wave, are located in the wavenumber plane
at the intersections of the parametrically defined line
\begin{subequations}
\begin{align}
k_x &= \kappa_0 \sin\theta + k_{x0}, \\
k_y &= -\kappa_0 \cos\theta + k_{y0},
\end{align}
\label{Eq: Kappa0Line}
\end{subequations}
with the level curve
\begin{equation}  \label{Eq: LevelCurve}
\mu(k_x,k_y)=\mu_0+\eta\epsilon^2.
\end{equation}
Existence of such intersections means that the line soliton is in
resonance with Bloch modes.

Inspection of the level curves in Fig. \ref{Fig: XSoliton} shows
that for almost all angles $\theta$, the line (\ref{Eq:
Kappa0Line}), whose slope is $-\cot\theta$, always intersects those
level curves for both $\eta=\pm 1$. The only exceptions are
$\theta=0, \pm \pi/4$ for $\eta=1$ and $\theta=\pi/2$ for $\eta=-1$.
In these cases, resonance is absent, thus true line solitons could
exist. Recalling the conditions for envelope solitons below Eq.
(\ref{Eq: aD}), we see that true line solitons with angles
$\theta=0, \pm \pi/4$ may bifurcate from the X-point (\ref{Eq:
Xpoint}) under defocusing nonlinearity ($\sigma=-1$), and true line
solitons with angle $\theta=\pi/2$ may bifurcate from this X-point
under focusing nonlinearity ($\sigma=1$). But where are the envelope
positions of these bifurcating line solitons inside Bloch bands? To
answer this question, it is again necessary to use exponential
asymptotics.

Strictly speaking, our exponential-asymptotics analysis in the
previous sections was for line solitons bifurcating from edges of
Bloch bands and residing outside of them (thus resonance with Bloch
bands never occurs). But for line solitons with special angles
inside Bloch bands, those which avoid resonance, our previous
analysis can be applied without any additional work. The conclusion
is that, for each of those special angles, two families of line
solitons, one onsite and the other offsite, bifurcate out from this
X-symmetry point; and their envelope locations are given by the
formulae (\ref{Eq: w0}). Numerically we have confirmed these
predictions. For instance, the onsite line solitons near the X-point
(\ref{Eq: Xpoint}), at angle $\theta=\pi/4$ for defocusing
nonlinearity and at angle $\theta=\pi/2$ for focusing nonlinearity,
have been numerically obtained and displayed in Fig. \ref{Fig:
XSoliton} (right panels); and their envelope locations agree with
those predicted by the formulae (\ref{Eq: w0}). Our analytical
results are consistent with the numerical and experimental reports
of line solitons inside Bloch bands in
\cite{Wang2007,Yang2011,PeliYang2012} as well.

\begin{figure}[htbp!]
\begin{center}
\includegraphics[width=2.25in]{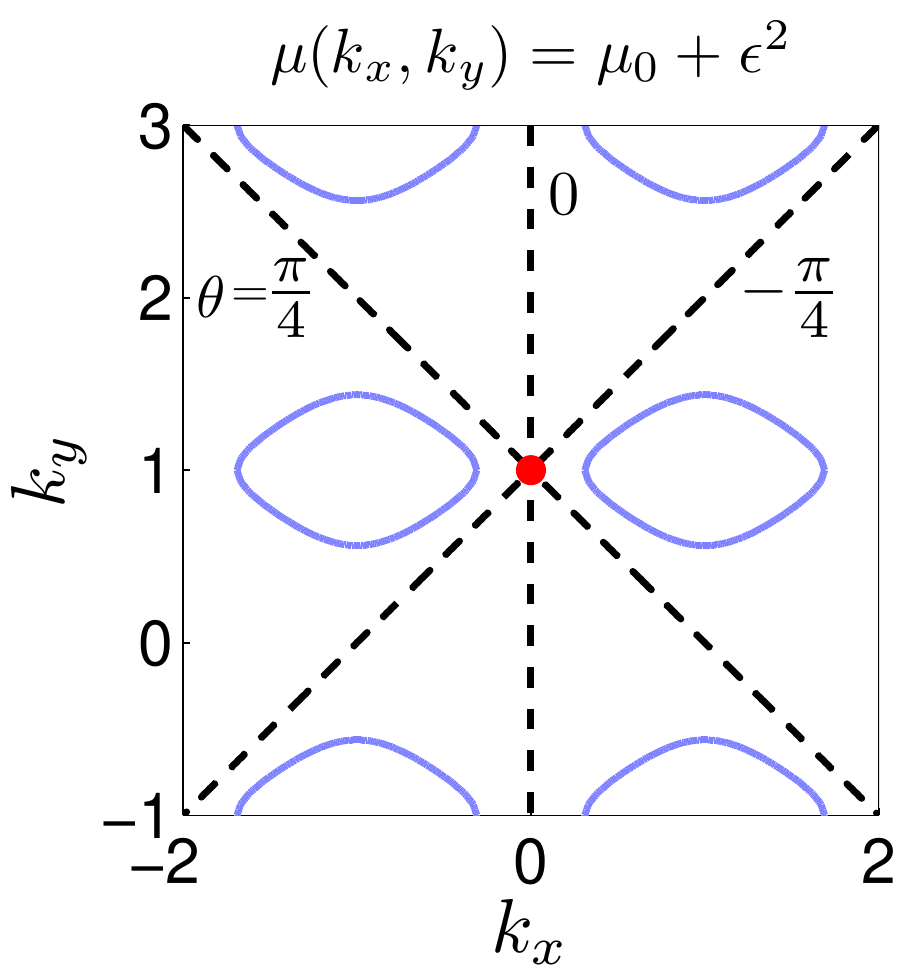}
\includegraphics[width=2.25in]{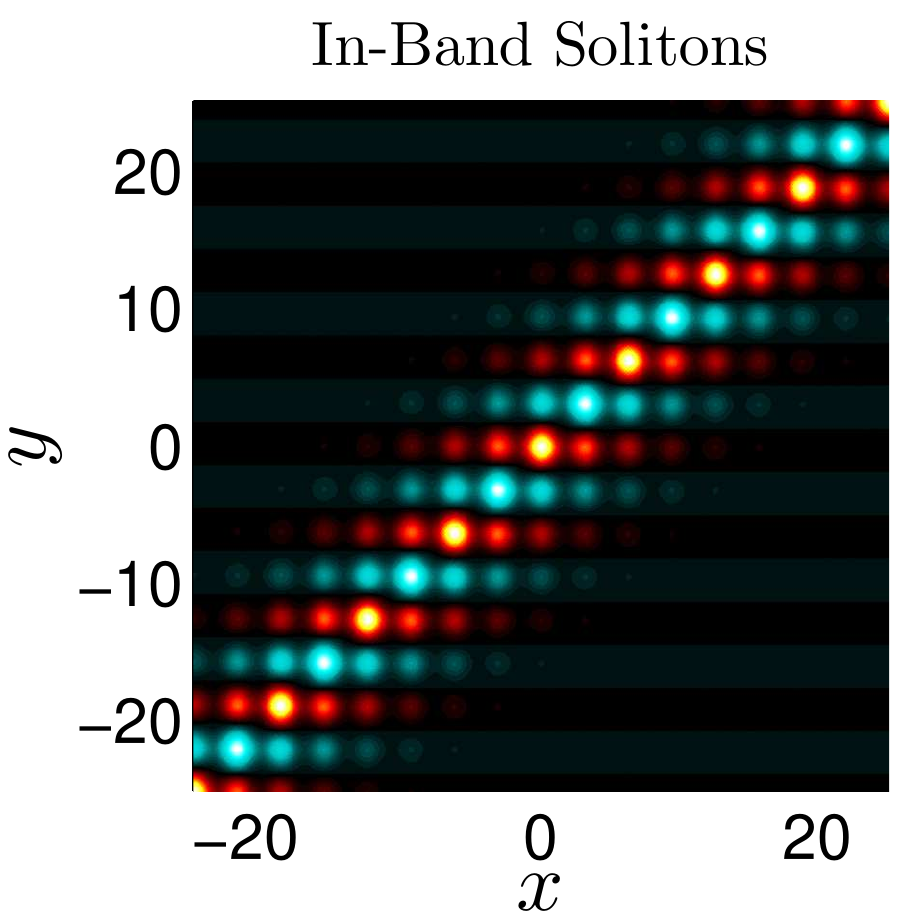}

\vspace{0.2cm}
\includegraphics[width=2.25in]{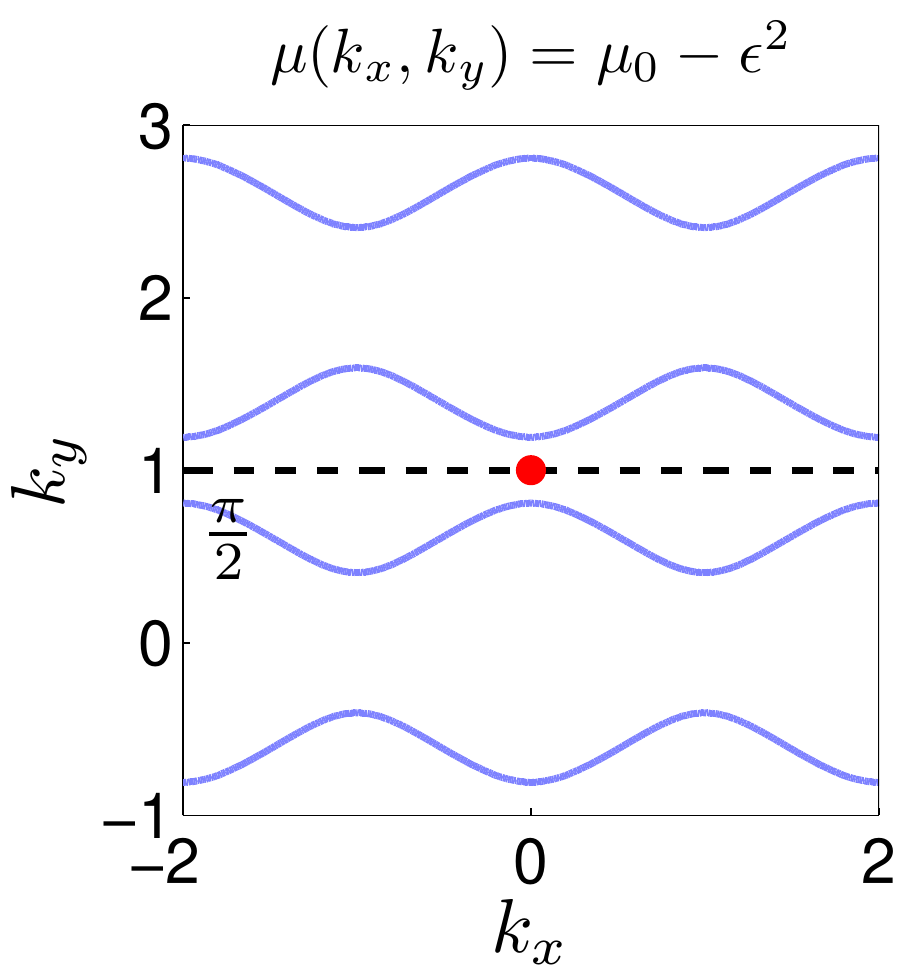}
\includegraphics[width=2.25in]{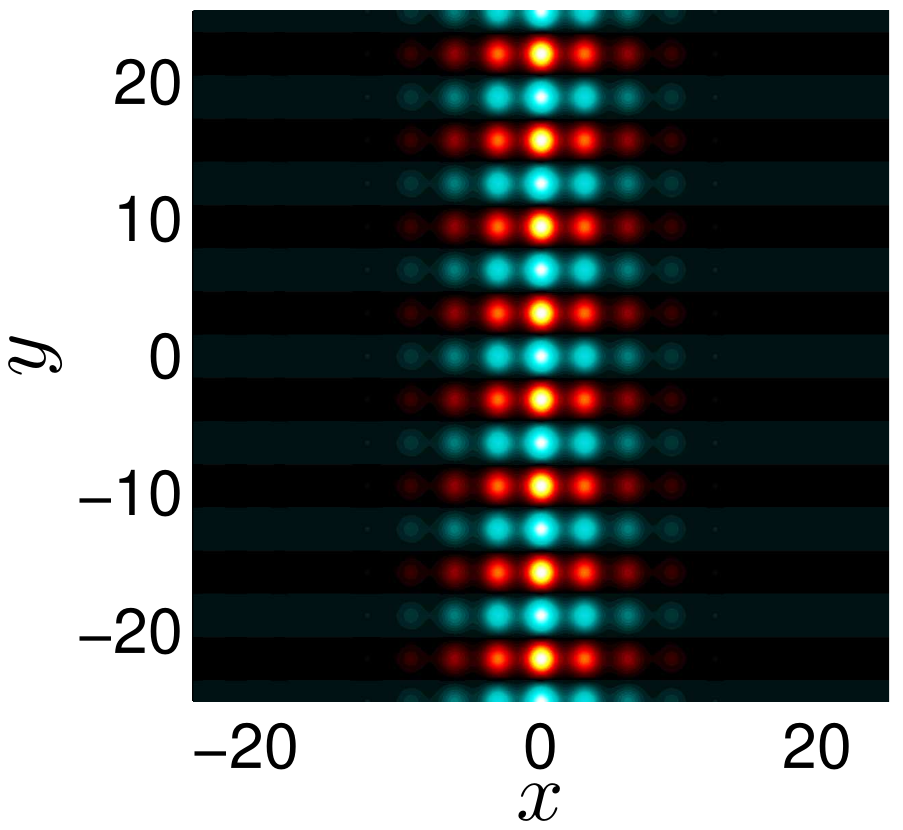}
\end{center}
\caption{(Color online) In-band line solitons in the lattice
\eqref{Eq: Xlattice} near the X-symmetry point $(k_x, k_y)=(0, 1)$
at special angles for $\epsilon = 0.2$. Top: $\eta=1$ and
$\sigma=-1$ (defocusing nonlinearity). Bottom: $\eta=-1$ and
$\sigma=1$ (focusing nonlinearity). Left: level curves $\mu(k_x,
k_y) = \mu_0+\eta \epsilon^2$ extended periodically (solid) and the
line \eqref{Eq: Kappa0Line} (dashed) for values of angle $\theta$
which admit true line solitons. Right: onsite line solitons at these
special angles with $\mu=\mu_0+\eta \epsilon^2$. \label{Fig:
XSoliton} }
\end{figure}

As explained above, inside Bloch bands, line solitons for most
inclination angles are nonlocal (in the normal direction), in the
sense that they comprise non-vanishing Bloch-wave tails far away
from the central line. Then a natural question is, how can we
determine the amplitudes of those non-vanishing Bloch-wave tails? It
turns out that this question can also be treated in the framework of
the exponential-asymptotics theory developed above. Specifically, we
should first recognize that the $\kappa_0$ value at the intersection
of the line (\ref{Eq: Kappa0Line}) and the level curve (\ref{Eq:
LevelCurve}) also satisfies equation (\ref{Eq: LinearU}) for
singularity locations (which can be easily checked). Thus this
$\kappa_0$ value of resonant Bloch modes is also a real pole in the
Fourier transform $\widehat{\psi}(x,y,K)$ of the nonlocal
line-soliton solution $\psi(x,y, W)$. Unlike the fourth-order poles
(\ref{Eq: realkappa0}) in our earlier analysis (see Eq. (\ref{Eq:
UAsymptotics})), this resonant pole is simple (i.e., first-order).
The residue of this resonant pole is also exponentially small in
$\epsilon$, and its exact value can be computed from the same outer
integral equation (\ref{Eq: Uorder1}) by the same numerical
algorithm (\ref{Eq: Iteration}) in the earlier text. From the
residue of this resonant pole and the inverse Fourier transform
(\ref{Eq: IFT}), the amplitudes of non-vanishing Bloch-wave tails in
the nonlocal line solitons will then be obtained. This problem of
calculating amplitudes of non-vanishing Bloch-wave tails in nonlocal
line solitons resembles that of calculating amplitudes of
continuous-wave tails in the fifth-order Korteweg–de Vries equation,
the third-order NLS equation and other related equations
\cite{Yang_SIAM,Pomeau1988,Akylas_Grimshaw1992,
Akylas_Yang1995,GrimshawJoshi1995,Grimshaw1995,CalvoAkylas1997}. But
our treatment of directly solving the outer integral equation for
the residues of resonant poles differs from those earlier
one-dimensional works, which hinge on the recurrence relation for
the coefficients of the Taylor expansion of the Fourier transform.

The results from the above specific example hold qualitatively for
all line solitons inside Bloch bands. Specifically, inside Bloch
bands true line solitons exist only for very few (up to three)
special angles due to the requirement of resonance suppression. For
each of those special angles, there exist two line solitons, one
onsite and the other offsite.

\section{Conclusion}

In this paper, we have presented what we believe is the first step
toward a fully two-dimensional asymptotic theory for the bifurcation
of solitons from infinitesimal continuous waves. For line solitons
bifurcating from infinitesimal Bloch waves in a general
two-dimensional periodic potential, an analytical theory utilizing
exponential asymptotics is developed. For this two-dimensional
problem, the previous approach of relying on a certain recurrence
relation to solve the outer integral equation is no longer viable
due to the presence of complex poles which are closer to the origin
than the real poles. Instead, we solved this outer equation directly
along the real line up to the real poles. This new approach not only
overcomes the recurrence-relation limitation, but also simplifies
the exponential-asymptotics process.

Using this modified technique, we showed that from every edge of the
Bloch bands, line solitons with any rational line slope bifurcate
out; and for each rational slope, two line-soliton families exist.
In addition, a countable set of multi-line-soliton bound states have
been constructed analytically. Furthermore, as a byproduct of this
exponential-asymptotics theory, a certain linear-stability
eigenvalue that bifurcates out of the origin at a band edge, is
analytically obtained. Line solitons bifurcating from interior
points of the Bloch bands were investigated as well, and it was
shown that such solitons exist (inside Bloch bands) only for a
couple of very special line angles due to resonance with the Bloch
bands. These analytical predictions were compared with numerical
results for symmetric as well as asymmetric potentials, and good
qualitative and quantitative agreement was obtained.

Throughout the analysis, the potential in Eq. (\ref{Eq: NLS}) is
taken to be in minimal-period orientation. If the potential is not
aligned along that minimal-period orientation, all our results would
still hold, except that the smallest of the real poles in Eq.
(\ref{Eq: realkappa0}) would not be those given by Eq. (\ref{Eq:
kappa0}), but rather be a certain multiple of those numbers.

In this article, we assumed that the potential in Eq. (\ref{Eq:
NLS}) has equal periods in $x$ and $y$. If the $x$- and $y$-periods
are not the same, say $\pi$ in $x$ and $\chi\pi$ in $y$, with $\chi$
being the ratio between the two periods, then we can introduce the
scaled $y$ variable $\hat{y}=y/\chi$. In this scaled variable, the
periods of the potential are $\pi$ in both $x$ and $\hat{y}$, hence
satisfying the assumption in this paper. Due to this $y$-scaling,
the Laplacian $\nabla^2$ in Eq. (\ref{Eq: NLS}) changes to
$\partial_{xx}+\chi^{-2}\partial_{\hat y\hat y}$. For this scaled
``Laplacian", all our analysis is still valid (except for very minor
modifications), because our analysis does not rely on the equal
coefficients in the Laplacian at all. Following this analysis, we
shall find that from every edge of the Bloch bands, line solitons
with slopes of any rational number divided by $\chi$ bifurcate out;
and for each of those slopes, two line-soliton families exist. We
shall also find that from a high-symmetry point inside Bloch bands,
line solitons at only a couple of special angles may bifurcate out.

Finally, we point out that the analysis in this paper is developed
for line solitons bifurcating from high-symmetry points of the Bloch
bands where the Bloch mode is unique (see Assumption 1). At certain
points of the Bloch bands, however, the Bloch modes are not unique
\cite{Shi_Yang}. In such cases, nonlinear interactions between
different Bloch modes would occur \cite{Shi_Yang}. To treat
line-soliton bifurcations from such Bloch-band points, the
exponential asymptotics analysis of this article would need to be
generalized. Such generalizations will be left for future studies.

\section*{Acknowledgment}
This work is supported in part by the Air Force Office of Scientific
Research under grant USAF 9550-12-1-0244.

\begin{center}
\section*{Appendix: \hspace{0.1cm} Line solitons in stripe lattices}
\end{center}

In this appendix, we consider line solitons in stripe lattices,
i.e., lattices which vary only in one direction $V(x,y) = V(x)$. We
shall show that line solitons in such lattices only have complex
poles and no real poles. In addition, the envelopes of these
solitons can be arbitrarily located. Our conclusion will be that
complex poles do not place restrictions on envelope locations.

The bulk of the analysis remains the same as in the main text. We
consider stationary solutions whose leading-order term is a
Bloch-wave packet,
\begin{equation}  \label{Eq: psi1D}
\psi(x,y,W) \sim \epsilon A(W-W_0) b(x).
\end{equation}
The packet envelope $A(W-W_0)$ has a sech-profile and varies only in
the direction of $W=\epsilon(x\sin\theta-y\cos\theta)$, and
$W_0=\epsilon x_0$ is the location of this envelope.

\begin{figure}[htbp!]
\begin{center}
\includegraphics[height=2.5in]{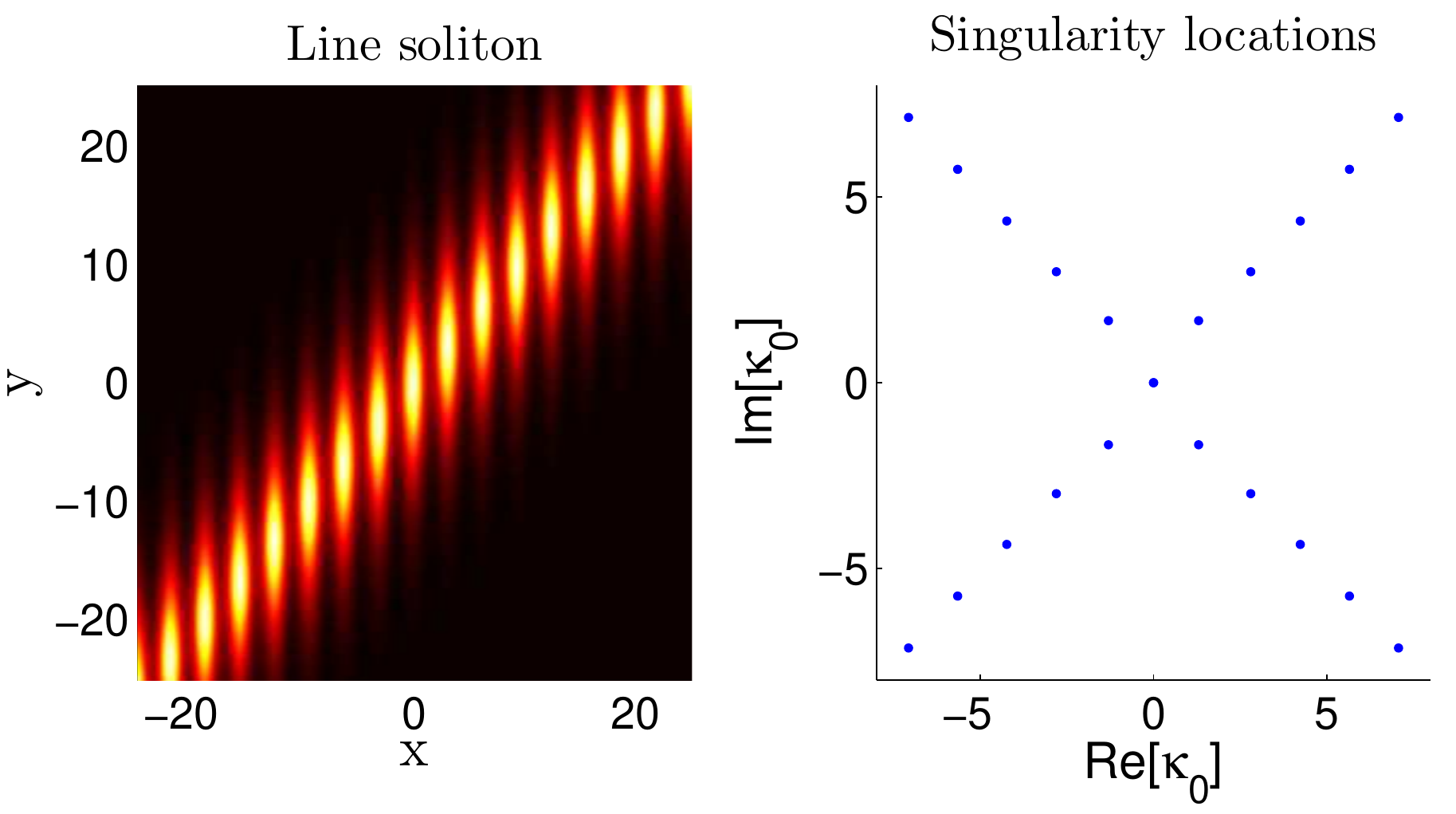}\\
\caption{(Color online) A line soliton (left) and its singularity
structure (right) for the stripe lattice $V(x)=6\sin^2
\hspace{-0.06cm} x$ with $\tan\theta = 1$ and $\sigma=1$.
\label{Fig: LStripe}}
\end{center}
\end{figure}

The pole singularities of these line wavepackets are the values of
$\kappa_0$ where Eq. (\ref{Eq: LinearU}) holds, except that the
operator $L_0$ drops the $y$-dependence now. That is,
\begin{equation}
L_0 \phi+ \ri \kappa_0 L_1 \phi - \kappa_0^2 \phi = 0,  \label{Eq:
EignoY}
\end{equation}
and
\begin{equation}
L_0 = \partial_x^2 + \mu_0 - V(x), \quad  L_1 = 2 \nabla \cdot
\left[ \sin\theta, -\cos\theta \right].
\end{equation}
Converting this equation into the eigenvalue problem \eqref{Eq:
EigenProblem}, we find that all eigenvalues $\kappa_0$ are now
complex-valued and not real (excluding $0$). This is illustrated in
figure \ref{Fig: LStripe} (right panel). Here the lattice is chosen
as $V(x)=6\sin^2x$ and the line slope as $\tan(\theta)=1$.

In this stripe lattice, we have numerically found line solitons. An
example with $\sigma=1$ is shown in figure \ref{Fig: LStripe} (left
panel). Since the lattice is $y$-independent, so are the Bloch modes
$b(x)$. Then if a line soliton exists with the envelope centered at
a particular $W_0$ value, from equation (\ref{Eq: psi1D}) it is
clear that line solitons with the envelope centered at arbitrary
$W_0$ values would exist, because any variation in $W_0$ may be
compensated for by a shift in the $y$ coordinate of $W$.

From the above analysis, we conclude that in a stripe lattice,
envelopes of line solitons can be arbitrarily positioned, and poles
of those solitons are all complex-valued (off the real axis).

\end{document}